\documentclass[twocolumn]{aastex62}
\usepackage{graphicx}
\usepackage{amsmath}
\usepackage{hyperref}
\usepackage{color}

\newcommand\bpmrp{G_{{\rm BP}}\,-\,G_{{\rm RP}}}

\newcommand\gaia{{\it Gaia}~}
\newcommand\panstarrs{{\it PanSTARRS}~}
\newcommand\rmi{$r_{\rm PS}\,-\,i_{\rm PS}$}
\newcommand\gmr{$g_{\rm PS}\,-\,r_{\rm PS}$}

\graphicspath{{./}{figures/}}

\accepted{July 9, 2020}
\published{August 14, 2020}
\submitjournal{ApJ}

\shorttitle{$531$ Halo White Dwarf Candidates from \gaia DR2}
\shortauthors{Kim, L{\'e}pine, \& Medan}

\begin{document}

\title{A Catalog of $531$ White Dwarf Candidates in the Local Galactic Halo from \gaia Data Release 2}

\correspondingauthor{Bokyoung Kim}
\email{bkim@astro.gsu.edu}

\author[0000-0002-8999-1108]{Bokyoung Kim}

\author[0000-0002-8999-1108]{Sebastien L{\'e}pine}

\author[0000-0003-3410-5794]{Ilija Medan}
\affiliation{Department of Physics and Astronomy, Georgia State University, 25 Park Place, Suite 605, Atlanta, GA 30303, USA}
 
\begin{abstract}
We present a catalog of $531$ white dwarf candidates that have large apparent transverse motions relative to the Sun ($v_{T} > 200$~km s$^{-1}$), thus making them likely members of the local Galactic halo population. The candidates were selected from the \gaia Data Release 2, and are located in a great circle with $20^\circ$ width running across both Galactic poles and Galactic center and anticenter, a zone that spans $17.3$\% of the sky. The selection used a combination of kinematic and photometric properties, derived primarily from \gaia proper motions, $G$ magnitudes, and $\bpmrp$ color, and including parallax whenever available. Additional validation of the white dwarf candidates is made using \panstarrs photometric ($gri$) data. Our final catalog includes not only stars having full kinematic and luminosity estimates from reliable \gaia parallax, but also stars with presently unreliable or no available \gaia parallax measurements. We argue that our method of selecting local halo objects with and without reliable parallax data leads us to round up all possible halo white dwarfs in the Gaia catalog (in that particular section of the sky) with recorded proper motions $> 40$~mas yr$^{-1}$, and that pass our $v_{T} > 200$~km s$^{-1}$~threshold requirement. We expect this catalog will be useful for the study of the white dwarf population of the local Galactic halo.
\end{abstract}


\section{Introduction} \label{sec:intro}
The identification of local stars from the Galactic halo population is an important tool to trace back the history of the Milky Way because halo stars are known to include the oldest stars of all the dynamical populations in the Galaxy \citep{carollo:16,kalirai:12}. Since the early 2000s, several big data survey operations successfully provided general schemes for the distribution and properties of Galactic halo stars \citep{juric:08,bond:10,lisanti:15}. However, those studies usually focus on the most luminous objects in the halo - like red giants and supergiants - because of the inevitable limitation of telescope capacities, even as we know that low-mass stars and white dwarfs must be by far the dominant objects in this very old population, and thus are the key to accurately mapping out and understanding the Galactic halo.

Due to their low intrinsic brightness, low-mass stars and white dwarfs in the halo remain generally out of range of current all-sky surveys, except in the solar vicinity ($d < 100 - 200$~pc). Within that relatively close range, halo stars are significantly outnumbered by Galactic disk stars, and their identification thus remains challenging. Their low spatial density also means that statistically significant samples of local halo objects must consist of stars that are significantly fainter on average than comparably large samples of disk stars from within the solar neighborhood ($d < 25 - 50$~pc).

In general, stars in the local halo population have higher spatial velocity relative to the Sun, compared to that of the disk population, which makes it possible to identify them as high proper motion stars. A case study for halo white dwarfs is the recent attempts to identify them among faint blue stars with large proper motions. These searches were motivated by the idea that white dwarfs may be a dark matter candidate because of their low absolute magnitudes but relatively high masses, prompting attempts to measure the local density of halo white dwarfs. After a search for high proper motion stars at high Galactic latitudes, \citet{oppenheimer:01} reported $38$ halo white dwarfs, which were suggested to represent the local population of halo white dwarfs. 

However, these results have been subject to debate. \citet{reid:01} argued that $75$~\% of the white dwarf candidates in \citet{oppenheimer:01} are not halo members but rather are part of thick disk populations with high rotational velocity. \citet{bergeron:05} also argued that most halo white dwarf candidates reported by \citet{oppenheimer:01} appear to be too warm and young to be part of halo populations unless their progenitors were low-mass main-sequence stars. In the wake of this debate, several attempts were made to identify true local halo white dwarfs based on more extreme kinematics \citep{lepine:05, eisenstein:06, kleinman:13, dame:16, munn:17}.

Since \citet{gaia:16, gaia:18b} released their second data set in early 2018, it has become possible to analyze detailed structures in the color-magnitude diagram (CMD). These astonishing results from \citet{gaia:18a} reveal unprecedented views of the color-luminosity distribution of nearby white dwarfs, and a catalog of $\sim70{\rm ,}000$ \gaia white dwarfs with accurate and reliable parallaxes was presented by \citet{jimenezesteban:18}. \citet{gentilefusillo:19} extracted a catalog of 260,000 high-likelihood white dwarf candidates from Gaia DR2 based on a selection in the HR diagram for stars with reliable parallaxes and with a likelihood probability, calculated from the relative distributions of a sample of spectroscopically confirmed white dwarfs from SDSS. In addition, \citet{kilic:19} have recently reported the identification of $142$ halo white dwarfs from \gaia DR2. They presented cooling ages for these white dwarfs based on a model atmosphere analysis using photometric and parallax data and complemented with a spectroscopic analysis of some objects. As a result, they argued that the age of the inner halo is consistent with the measurements of the ages of Galactic globular clusters. Because of the relatively small numbers found, one question is whether additional halo white dwarfs can be identified among the faintest objects in the \gaia catalog. \citet{brown:20} have also recently reported $98$ double white dwarf binaries from the Extremely Low Mass (ELM) survey, and they identified $22$ ELM white dwarfs in the Galactic halo on the basis of their $UVW$ velocities.

The surest way to unambiguously place a star in the halo is to have its full spatial motion, which requires an accurate measurement of the star's parallax and proper motions, and also its radial velocity. \gaia DR2 provides radial velocities for $\sim 7.2$ million stars, but only for relatively bright sources, which excludes most of the white dwarf candidates. Parallax and proper motion alone only provide a projection of a star's motion in the plane of the sky, but this can be used in specific cases to evaluation population membership, depending on the star's location on the sky, which determines the plane of projection. Local halo white dwarfs, however, tend to be too faint to even have reliable parallaxes in \gaia DR2. 

\citet{gaia:18b} reported a magnitude limit of $G \sim 19$ for accurate parallax determination. Therefore, parallaxes of stars below that magnitude limit will have unreliable projected velocity measurements at best. To make things worse, white dwarfs with unreliable parallaxes cannot be unambiguously identified from their location in the CMD. As it is highly risky to rely on only parallaxes for finding halo white dwarfs, proper motion measurements must take a more prominent role. Samples of stars with reliable \gaia proper motions can go more than $1$ or $2$ mag deeper than those with reliable parallax measurements. While at the faint end stars do not have precise \gaia parallaxes, it is still possible to use proper motion and magnitude information alone, under certain conditions, to identify halo white dwarfs and analyze their kinematics.

The goal of this paper is to compile the most extensive list of local halo white dwarf candidates from the local Galactic halo population, in an area covering $17.3$\% of the sky that is most appropriate to the identification of halo white dwarfs from proper motion data. In Section \ref{sec:methods}, we describe our algorithm for identifying halo white dwarf candidates from \gaia DR2. The list of candidates and further discussion about the white dwarf candidates in the Galactic local halo are described in Sections \ref{sec:result} and Section \ref{sec:panstarrs_setd}. We summarize our arguments and suggest possible future work in Section \ref{sec:summary}.

\section{Data and Methods} \label{sec:methods}
\begin{figure*}[!ht]
\centering
\plotone{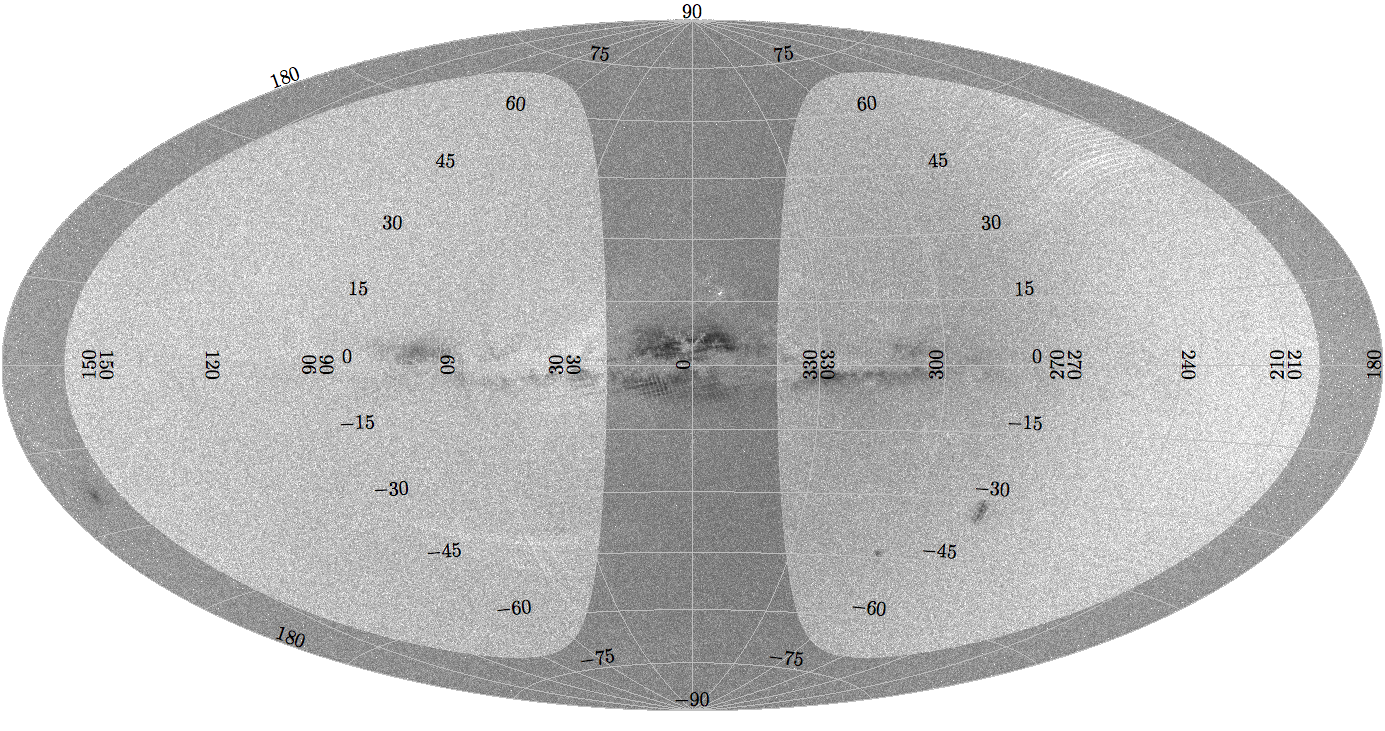}
\caption{Spatial distribution of the subset of $5.8$ million high proper motion stars in \gaia ($\mu_{tot} > 40$ mas yr$^{-1}$). The darker area shows the subsample selected for this study, which consists of all stars falling along the great circle with a width of $20^\circ$ passing across both Galactic poles and the Galactic center and anticenter. \label{fig:fig1}}
\end{figure*}

\subsection{\gaia DR2 Subset with High Proper Motions} \label{sec:gaiasubset}
We first extracted and assembled a \gaia subcatalog containing $\sim 5.8$ million stars with reported high proper motions ($\mu > 40$~mas yr$^{-1}$). As seen in Figure~\ref{fig:fig1}, those stars show a relatively uniform spatial distribution on the sky, consistent with a local stellar subset. Next, we elect to focus on stars located near a great circle with a width of $20^{\circ}$, crossing over both Galactic poles and the Galactic center and anticenter. About $1.8$ million stars are collected from the sample selection area, shown as the darker area in Figure~\ref{fig:fig1}. Choosing stars along this great circle is the first step in catching halo stars showing a large asymmetric drift because one of the proper motion components of the stars within this area of the sky runs parallel to the direction of the Sun's rotational motion in the Galaxy ($V_{\parallel}$). We will discuss this in \S \ref{sec:coordconv} in more detail.

\subsection{Coordinate Conversion}\label{sec:coordconv}
\gaia DR2 not only provides distance information but also gives us kinematic information from the combination of parallax and proper motion, yielding transverse velocities. These transverse velocities provide only partial (i.e. plane of the sky) kinematic information, however, and must be used with caution unless we use well-defined subsets in specific areas on the sky: one example is the use of a great-circle subset. With an appropriate choice of great circle, one can measure the component of motion of a set of stars in one particular $(U,\,V,\,W)$ direction in velocity space. This is because when you select stars along a great circle, one of the proper motion components for every stars - the component perpendicular to the great circle itself - is pointing in a specific direction. If you can calculate the proper motion component perpendicular to the circle, that component will thus be parallel to the same projected motion for all of the stars. For example, the transverse velocities in the Galactic longitude direction, $v_{T,l}$, of stars in the darker area of Figure~\ref{fig:fig1} generally point in the direction perpendicular to the great circle, which is in the parallel direction to the apex/antapex of the Sun's orbital motion in the Galaxy (component $V_{\parallel}$). On the other hand, transverse motions in the Galactic latitude direction, $v_{(T,b)}$, which point in a direction parallel to the great circle, can be interpreted as one of the components of motion running perpendicular to $V$, which we will denote $V_{\perp}$ but which represents a specific combination of the $U$ and $W$ components of motion, different for every star on the great circle.

The above description is, however, an oversimplification because the component of proper motion $\mu_{l}$ along the Galactic longitude is not in fact everywhere perpendicular to the great circle, especially for stars at high Galactic latitude ($b > 70^{\circ}$). To determine the component of proper motion that runs perpendicular to the great circle, we introduce a coordinate system $(r,\,s)$ that is tilted 90 degrees from the Galactic coordinate system. We simply convert all positions and proper motions using the following procedure: (1) We convert the positions and proper motions of all stars from 2D Galactic coordinates, ($l,\,b$), into Galactic Cartesian coordinates, $(x, y, z)$ (Equation~\ref{eq:eq1}). (2) We apply the rotation matrix to those position vectors (Equation~\ref{eq:eq2}), and then (3) we create the new coordinates ($r,\,s$) by restoring the position vectors of stars into the 2D sky grid (Equation~\ref{eq:eq3}):

\begin{equation}
\begin{split}
x_i &= \cos{l} \cos{b} \\
y_i &= \sin{l} \cos{b} \\
z_i &= \sin{b} \\
\\
\mu_{x_i} &= -\mu_{l}\sin{l}-\mu_{b}\sin{b}\cos{l}\\
\mu_{y_i} &= \mu_{l}\cos{l}-\mu_{b}\sin{l}\sin{b}\\
\mu_{z_i} &= \mu_{b}\cos{b}\\
\end{split}
\label{eq:eq1}
\end{equation}

\begin{equation}
\begin{split}
\begin{bmatrix}
x_f \\
y_f \\
z_f
\end{bmatrix}
&= 
\begin{bmatrix}
 1 & 0 & 0 \\ 
 0 & 0 & -1 \\ 
 0 & 1 & 0
\end{bmatrix}
\begin{bmatrix}
x_i \\
y_i \\
z_i
\end{bmatrix}
=
\begin{bmatrix}
x_i \\
-\,z_i \\
y_i
\end{bmatrix}
\\
\\
\begin{bmatrix}
\mu_{x_f} \\
\mu_{y_f} \\
\mu_{z_f}
\end{bmatrix}
&= 
\begin{bmatrix}
 1 & 0 & 0 \\ 
 0 & 0 & -1 \\ 
 0 & 1 & 0
\end{bmatrix}
\begin{bmatrix}
\mu_{x_i} \\
\mu_{y_i} \\
\mu_{z_i}
\end{bmatrix}
=
\begin{bmatrix}
\mu_{x_i} \\
-\,\mu_{z_i} \\
\mu_{y_i}
\end{bmatrix}
\end{split}
\label{eq:eq2}
\end{equation}

\begin{equation}
\begin{split}
r &= 
\begin{cases}
 y_f > 0.0,\,\,r = \frac{\arccos{x_{f}}}{\cos({\arcsin{z_f}})} \\
 y_f \le 0.0,\,\, r = 360^{\circ} - \frac{\arccos{x_{f}}}{\cos({\arcsin{z_f}})}
\end{cases}\\
s &= \arcsin{x_f} \\
\\
\mu_{r} &= -\mu_{x_f}\sin{r}+\mu_{y_f}\cos{r} \\
\mu_{s} &= -\mu_{x_f}\sin{s}\cos{r} - \mu_{y_f}\sin{r}\sin{s} + \mu_{z_f}\cos{s}  
\end{split}
\label{eq:eq3}
\end{equation}

(4) Finally, we can calculate the transverse motions from the general equation $v_{T}= 4.74 \times \mu \times d$ where $\mu$ is the proper motion in arcsec yr$^{-1}$ and $d$ is the distance (or distance estimate) in parsecs. The systematic error in the transverse velocity is $\mp 3.27$ km s$^{-1}$, assuming \gaia systematic proper motion errors of $\pm 0.02$ mas yr$^{-1}$ and a parallax zero point of $-0.029$ mas \citep{arenou:18, lindegren:18}. Note that stellar motions in the $s$ direction, for stars in the great-circle subset, approximately run parallel to the Galactic rotational velocities, $V_{\parallel}$, while those in the $r$ direction approximately correspond to a combination of $U$ and $W$ velocities, which we call $V_\perp$. Therefore, we calculate for each star $V_{\parallel} = 4.74 \times \mu_s \times d$, and $V_{\perp}= 4.74 \times \mu_{r} \times d$.

\subsection{Subsets for the White Dwarf Search}\label{sec:setcriteria}
The above discussion assumes that accurate proper motions and parallaxes are available, which is not always the case. Here we define two general subsets of different data quality that we call the {\it clean subset} (Set A) and the {\it unclean subset}; we further subdivide the  {\it unclean subset} into three different subsets (Sets B, C, and D). In this study, we neglect reddening corrections, which in normal cases can heavily affect colors and photometric distances, notably in the blue color regime typically used for white dwarfs. We believe reddening has relatively minor consequences for our particular subset because (1) the mean distance of our white dwarf candidates from \gaia parallaxes is $240$~pc, (2) most stars we selected are located above the Galactic plane, and (3) stars near the Galactic center ($|b| \le 20^{\circ}$) that can be severely affected by the reddening are mostly rejected from the selection (see \S \ref{sec:setbselection}). \citet{andrae:18} reported that the true reddening at high galactic latitude ($|b| > 50^{\circ}$) is almost near zero. Although the recent studies on 3D interstellar dust maps \citep{chen:19, lallement:19} reported the presence of some complex dust structures within 500 pc, those structures are mostly concentrated around near the Galactic center or anticenter within the low galactic latitudes.

\subsubsection{Set A: Clean Subset}
To assemble the {\it clean} subset (also called Set A), we partially applied Selection A, B, and C criteria introduced by the \gaia Collaboration. \citet{lindegren:18} guided \gaia users interested in assembling subsets of stars with reliable parallaxes through a general procedure for cleaning up stars with bad astrometric and photometric measurements. They recommended checking uncertainties in the parallax measurements, uncertainties in fluxes of the \gaia BP and RP filters, the renormalized unit weight errors (RUWE),\footnote{The renormalized unit weight error (RUWE)$= u/u_{0}(G, C)$ (Gaia Technical Note: GAIA-C3-TN-LU-LL-124-01, \url{https://www.cosmos.esa.int/web/gaia/dr2-known-issues})} and $\bpmrp$~color excess factors ($E$). The selection criteria we applied in this study to assemble Set A are the following:

\begin{enumerate}
\item $\pi > 0.0$
\item $-3.0 \le \bpmrp \le 6.0$
\item $10 \le G \le 21$
\item $\sigma(F_{{\rm BP}})/F_{{\rm BP}} \le 0.10$
\item $\sigma(F_{\rm {RP}})/F_{\rm {RP}} \le 0.10$
\item ${\rm RUWE} < 1.40$
\item $1.0 + 0.015 (G_{{\rm BP}}-G_{{\rm RP}})^2 < E < 1.3 + 0.06(G_{{\rm BP}}-G_{{\rm RP}})^2$
\end{enumerate}

We do not apply any cut based on parallax measurement errors, which means that some stars in our catalog could have large parallax uncertainties. This, however, has the advantage of not introducing a distance bias in our sample and also allows us to keep stars that may not have reliable parallaxes but that have precise proper motions. This approach is supported by \citet{gaia:18a}, who reported that most \gaia stars with $G \le 18$ have more reliable proper motion measurements compared to their parallax measurements. After using the above selection cuts, $1{\rm ,}255{\rm ,}151$ high proper motion stars are included in Set A.

\subsubsection{Set B: Unclean Subset with \gaia Parallaxes and $\bpmrp$~Colors}\label{sec:setbselection}
A limitation of using only Set A is that we lose $\sim 40$\% of all the high-proper motion stars in our initial sample. Although these rejected stars may have bad flux measurements, their proper motions are, in general, sufficiently reliable to be used for kinematic selection and analysis. Therefore, we build another subset from the stars that were filtered out in the Set A selection (above), but with the requirement that the star must have at least a positive \gaia parallax value and a reasonable $\bpmrp$ color value for a nearby star (i.e. colors in the range $-3.0 \le \bpmrp \le 6.0$).

The difference between Sets A and B is that stars in Set B potentially have large parallax and magnitude uncertainties, which may affect their distribution in the reduced proper motion (RPM) diagrams and CMDs (see right panels in Figure~\ref{fig:fig2}). Due to lower data quality, white dwarf candidates identified in Set B may include more false positives than those from Set A. In particular, there is a higher chance of false positives near the Galactic center, the most crowded region on the sky, where \gaia has notoriously been experiencing problems in accurately measuring astrometric and photometric parameters for field stars \citep{arenou:18, lindegren:18}. Thus, we are excluding stars near the Galactic center in Set B ($\left | b \right | \le 20^{\circ}$) in order to minimize contamination from photometrically bad sources, and this condition is applied for Sets C and D as well. After applying these cuts, our Set B contains $522{\rm ,}760$ stars in total.

\subsubsection{Set C: Unclean Subset with No \gaia Parallaxes}
Some \gaia stars cannot even be included in Set B because some of them have negative \gaia parallaxes, or others do not even have any parallax measurements reported in \gaia. These stars must, however, have reported \gaia proper motions (from our initial $\mu \ge 40$ mas yr$^{-1}$) and $\bpmrp$ color values, which are sufficient to place them in the RPM diagram (see Section \S \ref{sec:rpm_setc}) and identify white dwarfs. We call this group of stars without \gaia parallaxes ``Set C." Stars near the Galactic center are excluded from the subset for the same reason mentioned in Section \ref{sec:setbselection} above. This subset includes $42{\rm ,}764$ stars, all having relatively ``primitive" data in \gaia DR2, compared to stars from Sets A and B. These stars will hopefully get more complete and accurate measurements in later \gaia data releases.

\subsubsection{Set D: Unclean Subset with No \gaia Colors}
Our final subgroup, or Set D, contains $28{\rm ,}476$ stars, which have proper motions but no $\bpmrp$ color measurements reported in DR2. As for Sets B and C, this set again excludes stars near the Galactic center. Many stars in Set D can have a negative parallax value or have \gaia $G$ magnitude. Since they do not have \gaia color measurements, we cannot use \gaia photometric data to select white dwarf candidates. However, we still have the opportunity to identify white dwarfs in Set D by obtaining reliable color data from other photometric catalogs, like \panstarrs. We are going to revisit this subset at the end of this paper (See Section \ref{sec:panstarrs_setd}), first focusing on the identification of white dwarfs in Sets A, B, and C.

\subsection{Identification of White Dwarfs with \gaia Parallaxes (Sets A and B)}\label{sec:rpmdiagram}
We first use a RPM diagram, which is especially useful for the identification and classification of various local stellar populations like young disk, old disk, and halo; see, for example, \citet{lepine:05}. The reduced proper motion (labeled $H$) is interpreted as the combination of photometric and kinematic information, but is simply calculated from a star's apparent magnitude and proper motion, for example, 

\begin{equation}
H_G = G + 5 \log{\mu_{\rm tot}} + 5
\end{equation}

The reduced proper motion can be expressed in terms of the absolute magnitude ($M$) and transverse motion ($v_{T}$) of a star,

\begin{equation}
H_G = M_G + 5 \log{v_T} + 1.621 
\end{equation}

if $v_T$ is expressed in km s$^{-1}$. Therefore, stars with higher reduced proper motions must have higher transverse velocities if they have the same absolute magnitudes. Because of this characteristic, halo populations that usually have higher spatial velocities are clearly separated out from disk populations in the RPM diagram.

\begin{figure*}[!ht]
\centering
\includegraphics[scale=0.39]{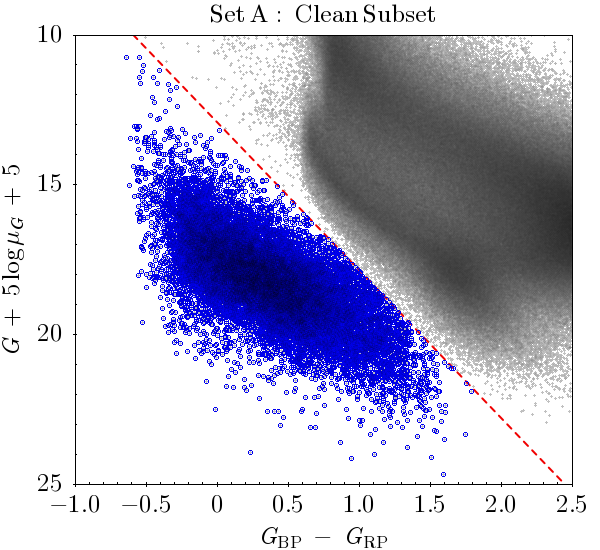}
\includegraphics[scale=0.39]{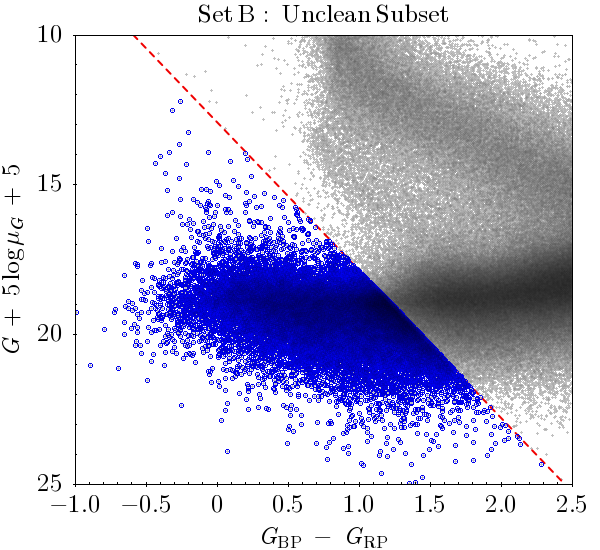}
\includegraphics[scale=0.39]{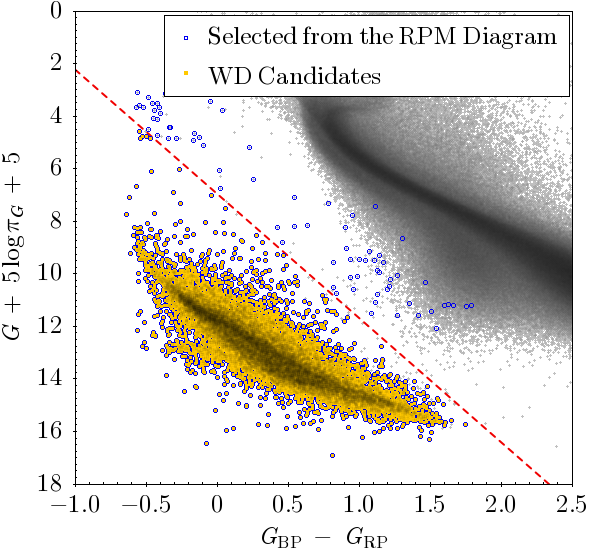}
\includegraphics[scale=0.39]{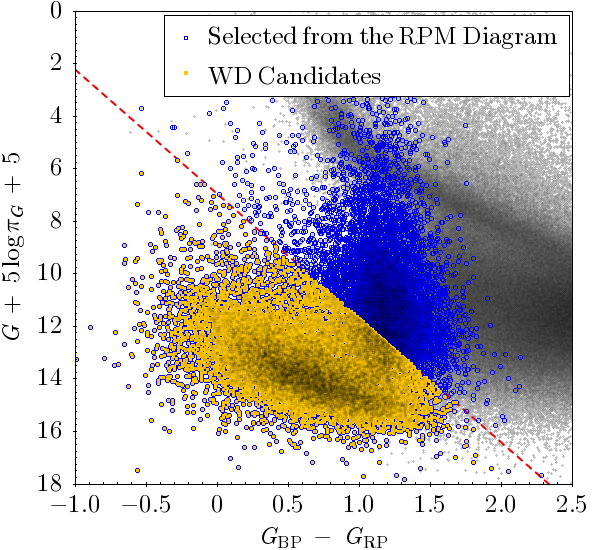}
\caption{RPM diagrams (top) and CMDs (bottom) for high proper motion stars from the clean subset (Set A; left) and for high proper motion stars from the unclean subset that have \gaia parallaxes (Set B; right). We defined an empirical color-RPM cut for the white dwarfs in the RPM diagram of Set A to select white dwarf candidates (blue points). Red-dashed lines in both RPM diagrams show our primary empirical cut between main-sequence stars and white dwarfs. We use a secondary cut to filter white dwarf candidates based on their distribution in the CMD (bottom panels), based on the locus of white dwarfs in the CMD of Set A. The finalized white dwarf candidates are shown as yellow points in each CMD. \label{fig:fig2}}
\end{figure*}

The RPM diagram was shown to be an especially useful tool in searching for nearby white dwarfs by \citet{limoges:13}. This is because white dwarfs have fainter absolute magnitudes compared to any other stellar populations in the Galactic disk or even halo. Therefore, the locus of white dwarfs is at the bottom left in the RPM diagram and is clearly distinct from the loci of the two main-sequence disk and halo populations, as for example in the upper panels in Figure~\ref{fig:fig2}, which shows the RPM diagrams of Sets A and B, respectively. In the diagram for Set A, we define an empirical color-RPM cut, shown as a red dashed line that efficiently separates the halo main-sequence stars and white dwarfs and follows the linear equation

\begin{equation}
[H_{G}]_{\rm lim} = 4.94 (G_{BP} - G_{RP}) +  12.91.
\label{eq:eq6}
\end{equation}

This cut is simply defined by the density distributions of halo main-sequence stars and white dwarfs in the CMD. We drew $\bpmrp$ histograms in each of $10$~reduced proper motion bins, obtained inflection points (minima) of the overall number distribution, selected the median $\bpmrp$ value where the number of stars is nearly zero, and then performed a 1D polynomial fit of the inflection points as a function of the reduced proper motion of the bin to get the linear line. Stars below the cut (blue points) are hence identified as probable white dwarfs.

To demonstrate the efficiency of the RPM diagram in identifying white dwarfs in Set A, we plot the distribution of the RPM-selected white dwarfs in the CMD (bottom left panel in Figure~\ref{fig:fig2}). The diagram shows that the overwhelming majority of RPM-selected objects are in the expected locus of white dwarfs, in the bottom left side of the plot. We only find a small number of RPM-selected objects that appear to be either hot subdwarfs or metal-poor, low-mass stars; all these objects may owe their low RPM values to unusually large transverse motions.

From the CMD distribution, we define an additional empirical cut between the main sequence and white dwarfs, which is drawn as a red dashed line, using the same method to get the empirical cut in the RPM diagram, which follows the linear equation
\begin{equation}
[M_{G}]_{\rm lim} = 4.727 (G_{BP} - G_{RP}) +  6.953.
\end{equation}

This additional restriction eliminates a small fraction of candidates in Set A; these stars are most likely to be either hot subdwarfs (sdO/sdB) or metal-poor, low-mass stars.

Set B, on the other hand, contains stars with less accurate photometric and astrometric data than Set A. The empirical lines defined with the stars from Set A eliminate a much larger number of stars in Set B. On close examination, the RPM diagram of Set B (top right panel in Figure~\ref{fig:fig2})  shows a large number of stars that are distributed horizontally along $H_{G} \sim 18-19$, with a significant vertical scatter. That is because Set B is dominated by faint high proper motion stars whose parallaxes tend to be less reliable, and these tend to be found at the bottom of the RPM diagram. There is no clear boundary between halo stars and white dwarfs in the RPM diagram for Set B, which suggests significant contamination from main-sequence, low-mass stars, very close to the red dashed line. This low-mass star contamination is in fact easy to identify in the CMD for Set B (bottom right panel in Figure~\ref{fig:fig2}). We see a locus on the bottom left that appears to be the spreading white dwarf sequence, but we also see another distinct clump above the white dwarf sequence and closer to the red dashed line. Stars in this clump are most likely low-mass stars with high transverse velocities but bad parallax measurements. Our additional cut in the CMD thus has the advantage of eliminating a large number of these contaminants, leaving a larger fraction of true white dwarfs in the subset. However, this also indicates that Set B may still suffer from some level of contamination, which will have to be taken into account.

Our initial samples of white dwarf candidates from Sets A and B are shown as yellow points in the CMD, and the number of stars are $17{\rm ,}692$ and $16{\rm ,}908$, respectively. These constitute our starting samples for identifying local halo white dwarfs (see Section \ref{sec:result}).

\subsection{Identification of White Dwarfs with No \gaia Parallaxes (Set C)}\label{sec:rpm_setc}

\begin{figure}
\includegraphics[scale=0.4]{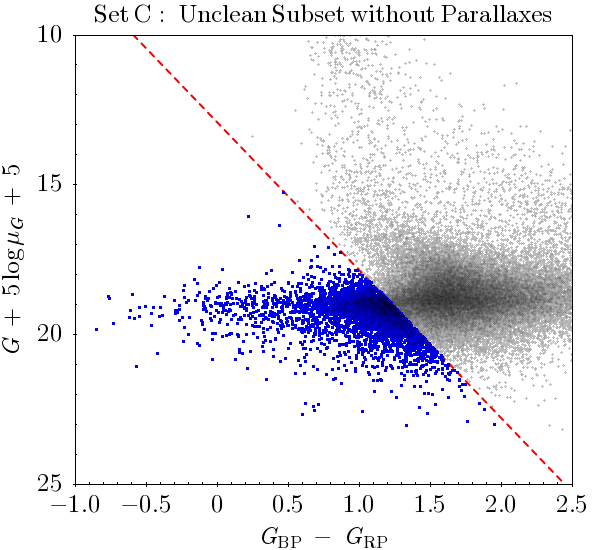}
\caption{RPM diagram for Set C. We used the same empirical color-RPM cut defined from Figure~\ref{fig:fig2} to select white dwarf candidates (blue points).\label{fig:fig3}}
\end{figure}

Figure~\ref{fig:fig3} shows the RPM diagram for the stars in Set C, the subset of stars with no \gaia parallaxes but with reliable proper motions. Blue points are primary white dwarf candidates selected by using the empirical cuts in the RPM diagram that were defined for Set A. The number of white dwarf candidates (blue points) is $4368$, but our experience with Set B (see above) suggests that a significant fraction of these may be main-sequence star contaminants. Indeed, a substantial fraction of the candidates are close to the selection line and are most likely to be main-sequence stars. These stars have very large reduced proper motions in any case and are most probably members of the halo population. The surest way to confirm whether or not they are actual white dwarfs is to verify that these objects have colors consistent with white dwarfs (and not main-sequence stars) in other photometric surveys. In Section \S \ref{sec:panstarrs_setc} below, we will describe how this can be done for stars in Set C, using their proper motion values and additional photometric information from \panstarrs Data Release 1.

\subsection{Collecting Additional Photometric Data from \panstarrs DR1}\label{sec:panstarrs}

\begin{figure}
\centering
\includegraphics[scale=0.39]{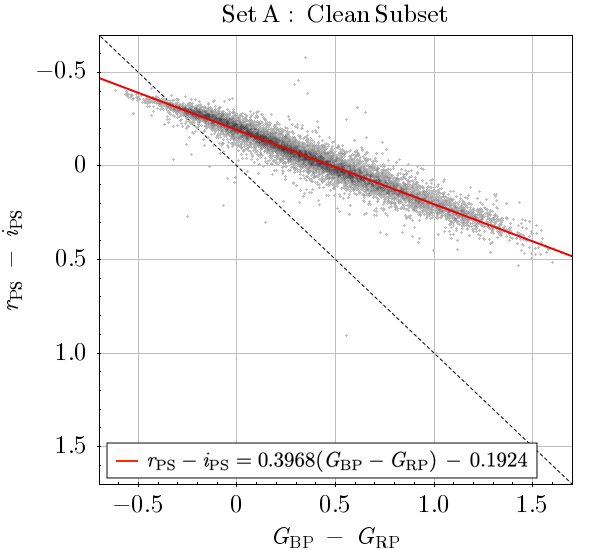}
\includegraphics[scale=0.39]{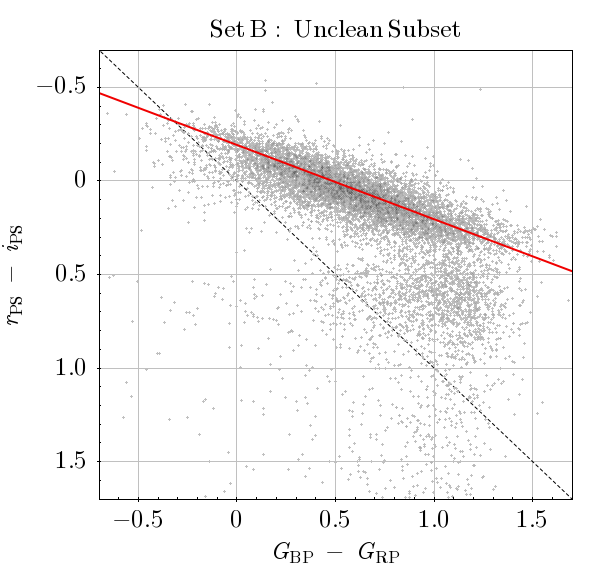}
\caption{Comparison of \panstarrs \rmi~color with $\bpmrp$ for white dwarf candidates from Sets A and B. The red line represents the best polynomial fit to the data in Set A. This relationship is only applicable in the \rmi~color range $-0.5 \le r_{\rm PS}\,-\,i_{\rm PS} \le 0.5$. Set B clearly shows more scatter ($\sigma = 0.30524$) in the relationship, consistent with larger measurement errors in $\bpmrp$. \label{fig:fig4}}
\end{figure}

To better characterize and vet our white dwarf candidates, we collected photometric information in {\it PanSTARRS} DR1 in order to get their \gmr~and \rmi~colors. The purpose of collecting \panstarrs data was to verify the consistency of our white dwarf candidates in the (\gmr, \rmi) color-color diagram, which does not rely on \gaia parallaxes or kinematics. From this test, we expect to independently confirm the reliability of our comprehensive search for white dwarf candidates in \gaia DR2.

The \gaia DR2 Archive tentatively provides various cross-match results with external catalogs,\citep{marrese:19}\footnote{\url{https://gea.esac.esa.int/archive/}} including a cross-match with \panstarrs DR1. There are three sets one can potentially use: {\it best-neighbor}, {\it good-neighborhood}, and {\it original\_valid} catalogs. Although it is easier to get best-match results from the {\it best-neighbor} catalog, there is a possibility that this catalog accidentally missed many of our local halo stars. This is mostly because those stars are hard to track due to their high proper motions. Therefore, we need to develop a new cross-match algorithm to recover the missing counterparts for the high proper motion objects.

We therefore conducted our own cross-match of {\it PanSTARRS} DR1 to our full set of high proper motion stars from \gaia DR2 using a Bayesian method (I. Medan \& S. L{\'e}pine 2020, in preparation). In this analysis, the motion-corrected position and brightness of a \gaia source are compared to the positions and brightnesses of the \panstarrs sources within $30"$, such that 2D distributions of magnitude difference (between \gaia and \panstarrs DR1) versus angular separation are formed for various cuts of Galactic latitude and \gaia $G$ magnitude. To determine Bayesian probabilities for true matches with our catalog, representative local distributions of field stars are created by displacing the \gaia sample by $\pm 2'$ (depending on if the source is in the northern or southern hemisphere) to create 2D distributions representative of random field stars where, due to the small shift in decl., the stellar density of field stars is statistically comparable to that of the true catalog that is being searched \citep{lepine:07}. Using the distributions from our cross-match and those for random field stars, Bayesian probabilities for \panstarrs DR1 sources that are a match to our high proper motion \gaia catalog objects were calculated. The sample we kept for this study consists of possible counterparts with a Bayesian probability $> 95\%$; we find that the number of matches is significantly larger than the number of matches provided by the internal \gaia cross-matches. As a final precaution, we filtered the sample by $gri$ filter saturation limits (see Table~\ref{tab:tab1}), and we also got rid of stars with unreliable error values in $gri$ magnitudes ($\sigma_{gri} < 9999$). Consequently, we recovered counterparts to $674{\rm ,}619$ stars in Set A, $311{\rm ,}891$ stars in Set B, $28{\rm ,}009$ stars in Set C, and $8172$ stars in Set D, all of which now have reliable \panstarrs colors.

To verify the reliability of \gaia colors, we examine the correlation between $\bpmrp$ color and \panstarrs $r_{\rm PS}\,-\,i_{\rm PS}$; the results are shown in Figure~\ref{fig:fig4}. The top and bottom panels show white dwarf candidates from Sets A and B, respectively. Overall, \panstarrs colors have tight correlations with \gaia $\bpmrp$, and a relationship can be derived from a polynomial fit in the white dwarf color range. Stars in Set B have larger flux errors in the \gaia BP and RP filters, and thus show more dispersion ($\sigma = 0.30524$) in the relationship than that in Set A ($\sigma = 0.21898$). We defined the best-fitting relationship between \rmi~color and $\bpmrp$ color in the white dwarf color range using stars in Set A (red line). These relationships are only applicable for stars in a certain \rmi~color range: $-0.5 \le r_{\rm PS}\,-\,i_{\rm PS} \le 0.5$.

\section{Results and Discussion} \label{sec:result}
\subsection{Selection of Halo White Dwarf Candidates from Sets A and B}\label{sec:setakinematics}

The spatial velocity of a star in the solar neighborhood can be used to distinguish halo stars from the disk population. \citet{bensby:14} suggested that the halo population is dominated by stars with total spatial velocities $V_{tot} \equiv \sqrt{U^2+V^2+W^2} > 200\,\, {\rm km \,\,s} ^{-1}$, which normally requires one to know the full 3D space motion of a star from combined proper motion, parallax, and radial velocity measurements. A more flexible standard \citep{gaia:18a} is to use transverse velocity alone and assume that a star with $v_{T} > 200 \,\,{\rm km \,\,s}^{-1}$ is a likely member of the local halo population; this criterion does not require one to know the star's radial velocity. 

Although we do have radial velocity measurements for some of our stars, we adopt a transverse-velocity-only criterion to identify halo members in our entire sample. However, our particular sky selection allows us to make a somewhat more reliable selection of halo stars. As explained in \S \ref{sec:coordconv}, transverse velocities in the $(r,\,s)$ coordinate system can be interpreted as representing $V_{\parallel}$ on the one hand (for the component of motion along the alternative sky coordinate $s$), and a combination of $U$ and $W$ on the other hand (for the component of motion along the alternate sky coordinate $r$):
\begin{equation}
\begin{split}
&V_{\perp}\,\approx\,U sin(r) + W cos(r) \\
&V_{\parallel}\,\approx\,V
\end{split}
\end{equation}

This is useful because it is the $V$ component that is the best diagnostic to tell if a star is a member of the disk or halo, as it directly relates to the asymmetric drift, which is the principal characteristic of the halo population.

\begin{figure*}
\centering
\includegraphics[scale=0.39]{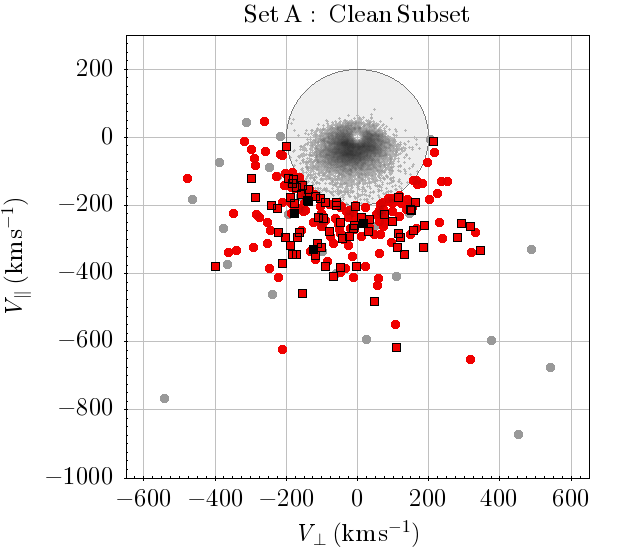}
\includegraphics[scale=0.39]{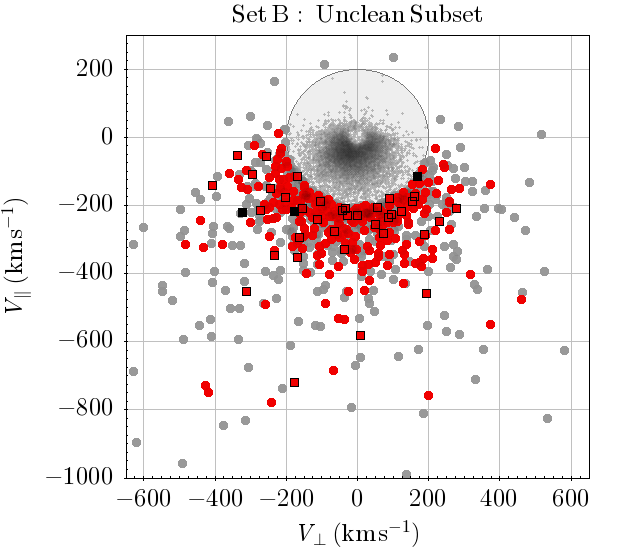}
\includegraphics[scale=0.39]{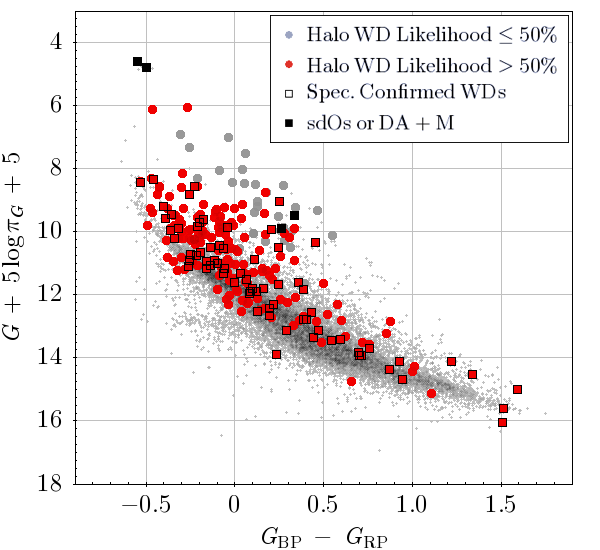}
\includegraphics[scale=0.39]{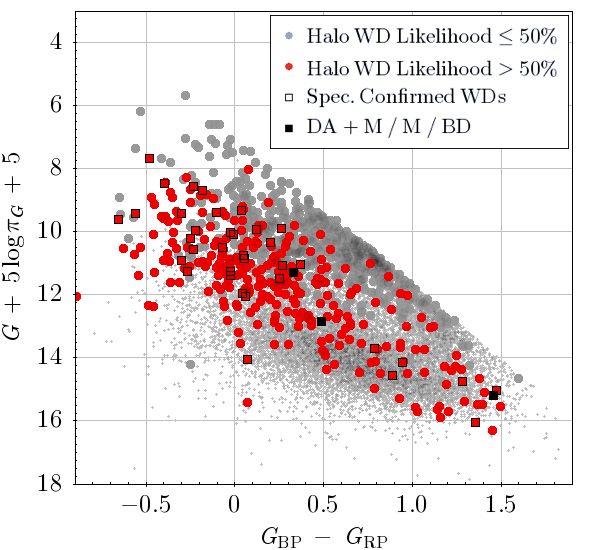}
\caption{Top panels: kinematic distributions of the halo white dwarf candidates in Sets A and B in the projected $(V_{\perp}, V_{\parallel})$ transverse velocity plane (see text). The inner circle is centered on the approximate location of the local standard of rest, defined by the local disk population. Small gray points in the inner circle show the distribution of all white dwarf candidates selected in the CMD, most of which appear to belong to the Galactic disk population. Large filled circles are white dwarf candidates with $v_{T} > 200$~km s$^{-1}$; red symbols are candidates with estimated likelihood $> 50$\%, while gray symbols are low-probability candidates. Black open squares are white dwarfs that have been confirmed spectroscopically in previous studies (see Tables~\ref{tab:tab3} and \ref{tab:tab4}), and black filled squares are objects known to be hot subdwarfs or WD+M binaries. Bottom panels: CMDs of the white dwarf candidates in both sets, showing the high-probability halo white dwarf candidates (red) falling near the expected white dwarf locus.
\label{fig:fig5}}
\end{figure*}

The top panels in Figure~\ref{fig:fig5} show the kinematic distributions of white dwarf candidates, selected from Sets A and B in the projected ($V_{\perp},\,V_{\parallel}$) plane. Gray points in the inner circle are white dwarf candidates with slow transverse velocities, $v_{T} \le 200$~km s$^{-1}$, which are most likely members of the disk population. Filled circles are the halo white dwarf candidates with high transverse motions, $v_{T} > 200$~km s$^{-1}$. We identify $249$ stars in Set A and $865$ stars in Set B that meet this criterion. There are $137$ stars in Set A and $447$ stars in Set B that have velocities $V_{\parallel} > -220$ km s$^{-1}$, which can be interpreted as having counterrotating orbits in the Galaxy relative to the local disk population. 

This possible contamination raises questions about confidence levels from this selection. To answer this, we ran simulations where we propagated random errors on the observed parameters of each of our candidate halo white dwarfs. We created possible observed values of the reduced proper motion, absolute $G$ magnitude, and total transverse velocity of each individual halo white dwarf candidate using a random number generator from the normal distribution, assuming each value observed from \gaia to be the mean value and each reported error to be the variance. We then evaluated for each simulated set whether or not the star would pass all of the selection cuts (RPM-color, $M_{G}$-color, and $v_{T}$ cuts) we defined above. If the star passed all three cuts, then we assign ``$1$", otherwise we assigned ``$0$". We ran 10,000 simulations for each star and from this calculated the likelihood of the star would pass the halo selection, and we reject stars with likelihood $\le 50$\% from the candidate list (gray filled circles in Figure~\ref{fig:fig5}). As a result, $218$ stars in Set A and $301$ stars in Set B remained as halo white dwarf candidates, and these are shown as red filled circles in Figure~\ref{fig:fig5}. All likelihood values expressed in percentages are listed in Table~\ref{tab:tab3}-\ref{tab:tab6}; we recommend checking these likelihood values before pursuing any further studies on the candidates.

A search of the astronomical literature determines that $74$ white dwarf candidates from Set A and $39$ white dwarf candidates from Set B were previously reported in various studies, including the halo white dwarf search by \citet{oppenheimer:01}, identifications of white dwarfs in SDSS DR7 \citep{eisenstein:06,kleinman:13} and in SDSS DR10 \citep{kepler:15}, and halo white dwarf searches by \citet{kilic:19}. Previously known white dwarfs are plotted in Figure~\ref{fig:fig5} as black open squares. Interestingly, nine objects in Sets A and B, shown as black filled squares, were previously reported to be either brown dwarfs \citep{zhang:19}, hot subdwarfs \citep{feige:58,green:86}, low-mass stars or binaries \citep{west:11}, or WD+M binaries \citep{eisenstein:06,li:14,rebassa:16}. Those are also included in the catalog and flagged appropriately.

The bottom panels in Figure~\ref{fig:fig5} show the distribution of the candidates from each set in the CMD, where both disk and halo stars are labeled as in the upper panels. The general distribution of white dwarf candidates from Set A (gray points) is consistent with the three white dwarf cooling sequences (A, B, and Q concentrations) introduced in \citet{gaia:18a}. The overall distribution from Set B is, however, more scattered and shows the dense clump above the white dwarf cooling sequence, as mentioned in \S \ref{sec:rpmdiagram}. We believe most stars in this clump are contaminating low-mass stars with bad parallax measurements. This idea is supported by the fact that all low-likelihood candidates (gray filled circles) are distributed in the upper clump.

\begin{figure}
\centering
\includegraphics[scale=0.39]{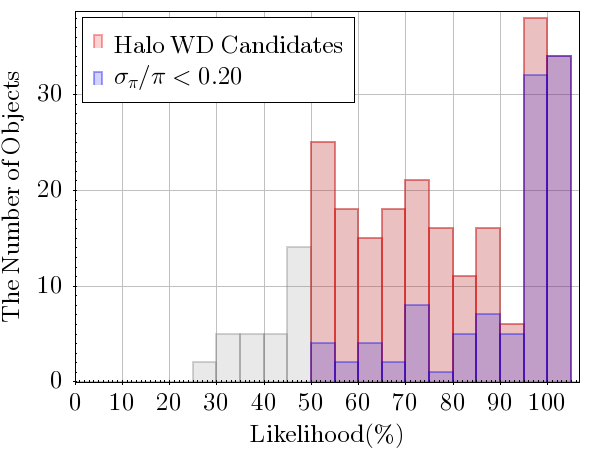}
\includegraphics[scale=0.39]{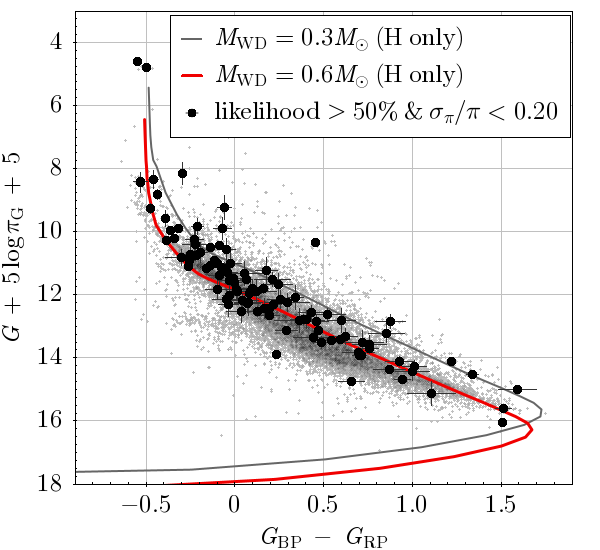}
\caption{Top: Likelihood distribution of stars with high transverse velocities in Set A. The red histogram shows the likelihood distribution of halo white dwarf candidates, and the blue histogram represents the distribution of the candidates with more precise \gaia parallax measurements. Bottom: CMD for the white dwarf candidates in Set A having accurate parallax measurements. The gray and red lines show the theoretical $0.3 M_{\odot}$ and $0.6 M_{\odot}$ hydrogen white dwarf cooling sequences. \label{fig:fig6}}
\end{figure}

The top panel in Figure~\ref{fig:fig6} shows the likelihood distribution of white dwarf candidates with high transverse velocities in Set A. The red histogram is the total distribution, while the blue histogram shows the distribution of candidates with more precise parallax measurements ($\sigma_{\pi}/\pi < 0.20$). This shows that the largest uncertainty of selecting halo candidates is coming from \gaia parallax measurements.

If we exclude stars with large errors and plot only stars with the most reliable astrometric measurements in the CMD (bottom panel in Figure~\ref{fig:fig6}), most of the contaminating populations are gone, and most candidates are distributed along the normal white dwarf sequences. Lines in the CMD show the cooling sequences of $0.3 M_{\odot}$ (gray line) and $0.6 M_{\odot}$ (red line) white dwarfs for pure hydrogen atmospheres \citep{holberg:06, kowalski:06, bergeron:11, tremblay:11}.\footnote{\url{https://www.astro.umontreal.ca/~bergeron/CoolingModels/}} Most candidates in Set A follow the $0.6 M_{\odot}$ cooling sequence, but about $10$\% of candidates are in better agreement with the low-mass white dwarf cooling sequence, raising the possibility that these may be old, low-mass white dwarfs, which are expected to appear more luminous due to their large radii. However, these stars would be more likely to be in interacting binary systems because it is not possible yet to form such low-mass white dwarfs through single-star evolution. Current stellar evolutionary theory predicts that a low-mass white dwarfs with a mass less than $\sim0.45\,M_{\odot}$ cannot be formed by the single-star evolution channel, but are most likely the result of a mass transfer event from the companion \citep{bergeron:92, kilic:07, pelisoli:19}.

In all likelihood, the few overluminous white dwarfs that still remain in the diagram are binary systems. Assuming a binary system composed of two white dwarfs of similar masses, it would only be overluminous by $0.7$ mag, which is not enough to explain many of our candidates in the diagram. However, the apparent overluminosity can be explained if the object is a WD + low-mass star (K or M dwarf) binary system, which should be moderately brighter but also significantly redder due to its low-mass companion star.

\subsection{Selection of Local Halo WD Candidates in Set B from Photometric Distances}\label{sec:photdist_setb}

One concern from the likelihood analysis in \S \ref{sec:setakinematics} is that the candidates in Set B may still include some contaminants that are due to their large astrometric errors. Although one might consider that having a parallax measurement is always an improvement over only having a photometric distance estimate, this is merely true only if parallaxes are measured in sufficiently high precision. In Set B, more than $94$\% of stars with high transverse velocities have such high fractional parallax errors ($\sigma_{\pi}/\pi > 0.20$). 

As an alternative to clearing Set B of contaminants, we generate the kinematic plot using photometric distances. Assuming our candidates are $0.6M_{\odot}$~white dwarfs with pure hydrogen atmospheres, we infer their absolute magnitudes from their $\bpmrp$~color, based on the the cooling sequence of \citet{bergeron:11}. Due to the dramatic changes at the blue and red ends of the cooling sequence, we only use candidates within the color range $-0.45 \le \bpmrp \le 1.60$; this includes the vast majority ($\sim 95$\%) of candidates, with only $14$ very blue/red stars excluded from the process.

\begin{figure}
\centering
\includegraphics[scale=0.39]{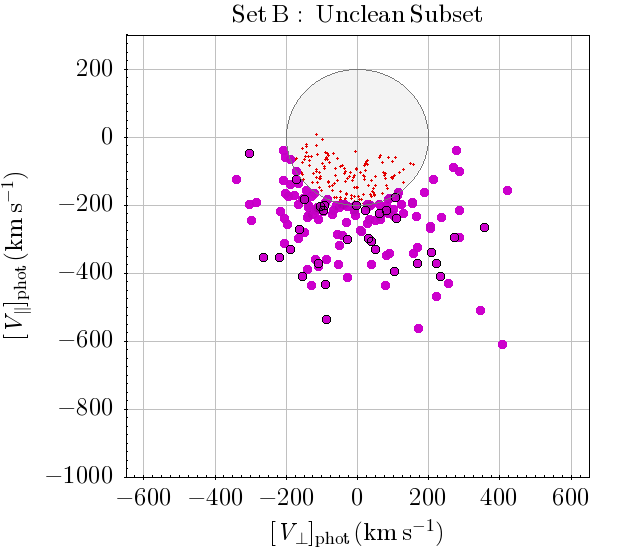}
\includegraphics[scale=0.39]{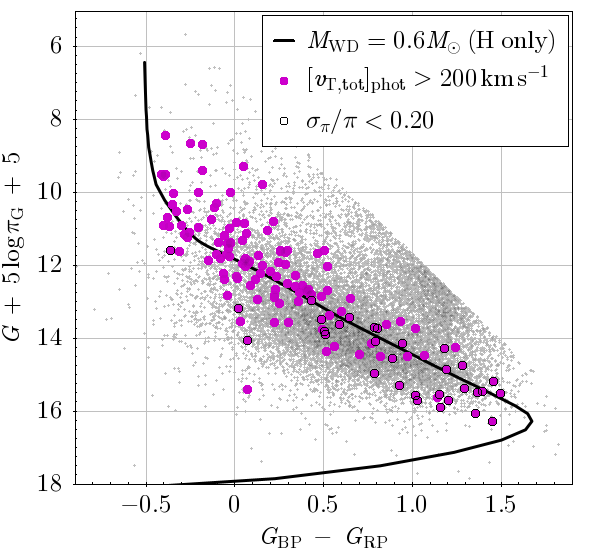}
\caption{Top: kinematic distribution of local halo white dwarf candidates in Set B with alternate values of calculated transverse velocities in the $r$ and $s$ directions based on their {\em photometric} distance. Red crosses are initial halo candidates now found to be low-transverse-velocity stars with this test. Purple filled circles ($141$ stars) are stars that are again found to have transverse velocities higher than $200$~km s$^{-1}$. Black open circles are halo candidates with precise parallaxes ($\sigma_{\pi}/\pi \le 0.20$). Bottom: CMD of the revised subset of halo white dwarf candidates selected from the diagram above. Black solid line shows the white dwarf cooling sequence we applied to calculate photometric distances to the candidates. The revised subset, though much smaller, shows a distribution in the CMD that is more consistent with the expected white dwarf locus, compared with the distribution in the bottom panel of Figure~\ref{fig:fig5}. \label{fig:fig7}}
\end{figure}

The top panel in Figure~\ref{fig:fig7} shows the result in the kinematic plane for the motions based on photometric distances. The plot is similar to that in Figure~\ref{fig:fig5}, but only shows the stars in Set B that were selected as likely ($> 50$\%) halo white dwarf candidates from their parallax-calculated space motions. If we now use the photometric-calculated space motions, we find that $51.0$\% of the stars ($146$~objects) now have kinematics more consistent with the disk. The rest of the stars ($141$~objects), on the other hand, still have kinematics consistent with the halo. The bottom panel in Figure~\ref{fig:fig7} shows the CMD of the $141$ reconfirmed halo candidates, shown as purple filled circles. Gray points are the full set of white dwarf candidates selected from the CMD in Figure~\ref{fig:fig2}, including the stars now rejected as being nonhalo white dwarf candidates based on their photometric distance estimates. Black open circles are halo candidates with precise parallaxes ($\sigma_{\pi}/\pi \le 0.20$). Most of our remaining halo white dwarf candidates appear to follow the theoretical $0.6M_{\odot}$~white dwarf cooling sequence, shown as a black solid line.

\subsection{Confirmation of Candidates from Sets A and B using {\it PanSTARRS} DR1 colors}\label{sec:panstarrs_sets_aandb}
\begin{figure*}[!ht]
\centering
\includegraphics[scale=0.39]{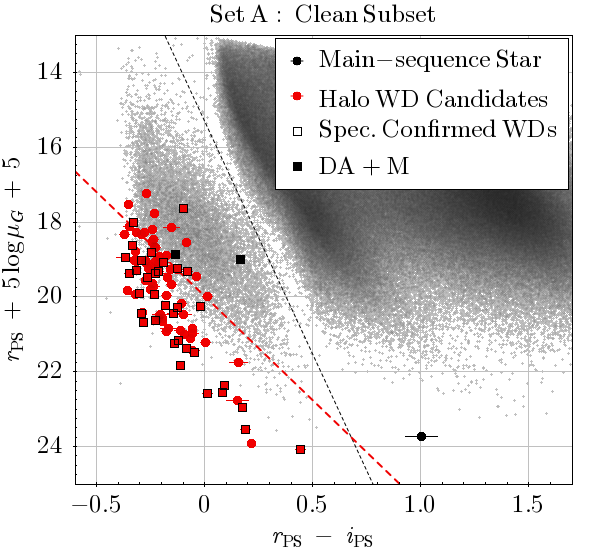}
\includegraphics[scale=0.39]{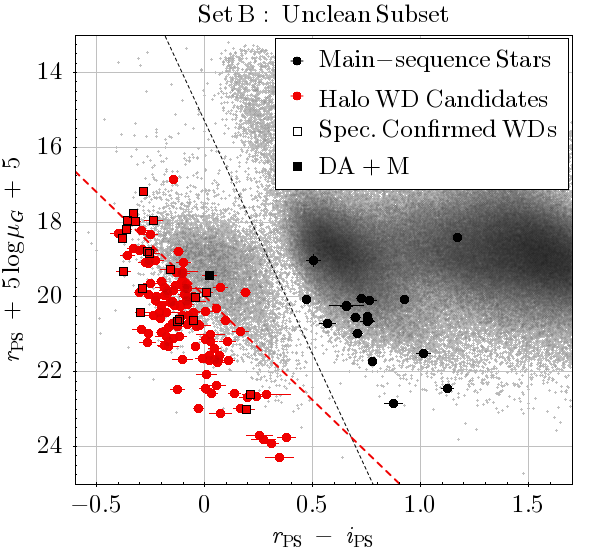}
\includegraphics[scale=0.39]{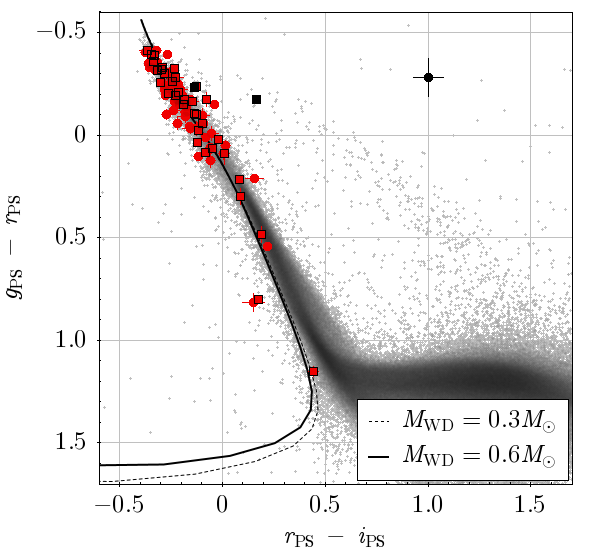}
\includegraphics[scale=0.39]{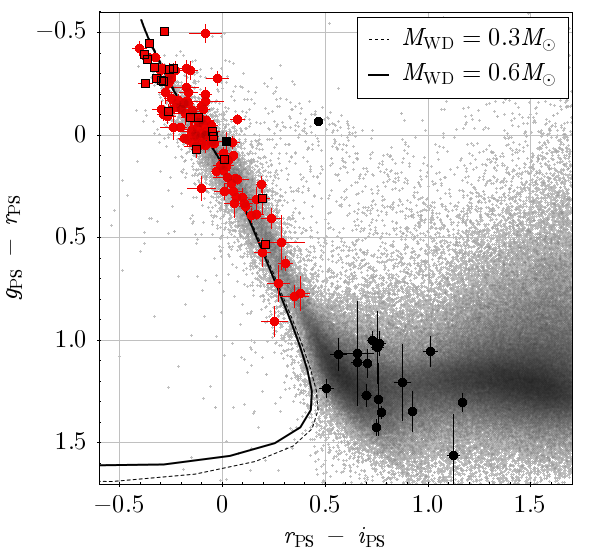}
\caption{Top panels: reduced proper motion diagrams of stars in Sets A and B as a function of \rmi~color. Each symbol in this diagram has the same definition as Figure~\ref{fig:fig5} (see legend). The black dashed line is the converted empirical cut based on the same cut we defined in \S \ref{sec:rpmdiagram} in \gaia color. Black filled circles in the diagram for Set B indicate stars having redder colors with respect to the black dashed line, which means they are probably not white dwarfs. The red dashed line indicates the boundary between normal white dwarfs and local halo white dwarfs, defined by shifting a linear fit of the white dwarf sequence in the RPM diagram (see text). Error bars show errors in reduced proper motion and \rmi~color. Bottom panels: \gmr~color distribution of stars as a function of \rmi~color. Black lines show the cooling sequences of white dwarfs for pure hydrogen atmospheres: $0.3M_{\odot}$ (dashed line), $0.6M_{\odot}$ (solid line), $1.0M_{\odot}$ (dashed-dotted line). Error bars indicate errors in \gmr and \rmi~color. \label{fig:fig8}}
\end{figure*}

The upper panels in Figure~\ref{fig:fig8} show RPM diagrams of stars in Sets A and B, this time using the \rmi~color from \panstarrs with the reduced proper motion $H_{r}$. Gray points represent all stars in Sets A and B with \panstarrs counterparts. These RPM diagrams show a clear segregation between main-sequence stars and white dwarfs, which allows us to define a clean empirical border, shown as a black dashed line, which is the analog of the separation line we defined in Figure~\ref{fig:fig5}: 

\begin{equation}
\begin{split}
[H_{r_{\rm PS}}]_{\rm lim} = 12.45(r_{\rm PS}\,-\,i_{\rm PS}) + 15.31 \\
(-0.5 \le r_{\rm PS}\,- \,i_{\rm PS} \le 0.5)
\end{split}
\label{eq:eq9}
\end{equation}

To define this limit, we converted the relationship between the reduced proper motion as a function of \gaia $G$ magnitude and $\bpmrp$, using the color conversion between $\bpmrp$ color and \rmi~color defined in \S \ref{sec:panstarrs}. In the diagrams, we identify a star in Set A and $18$ stars in Set B, shown as black filled circles, identified as white dwarf candidates in our initial selection, but now revealed in \panstarrs to have redder colors, consistent with the main-sequence population.

Stars to the blue of the black dashed line are, however, confirmed to be white dwarf candidates and are plotted in red filled circles in Figure~\ref{fig:fig8}; these consist of $98$ stars from Set A and $130$ stars from Set B. These stars line up along a distinctive locus consistent with a downshifted (i.e. high velocity) white dwarf cooling sequence. As in Figure~\ref{fig:fig5}, we plot white dwarfs and binaries confirmed in previous studies as black open or filled squares. A red dashed line is defined by the parallel shift of a linear fit of the standard white dwarf sequence in the CMD, and the magnitude of the shift is set by the distribution of our candidates in Set A so that more stars with likelihood $> 70$\% fall below the line:

\begin{equation}
\begin{split}
[H_{r_{\rm PS}}]_{\rm lim} = 5.56(r_{\rm PS}\,-\,i_{\rm PS}) + 19.98 \\
\end{split}
\label{eq:eq10}
\end{equation}

This line will guide our selection of halo white dwarf candidates in Sets C and D in Sections \ref{sec:panstarrs_setc} and \ref{sec:panstarrs_setd}.

The lower panels in Figure~\ref{fig:fig8} examine the distribution of the candidates in the (\gmr, \rmi) color-color plane with white dwarf cooling sequences (dashed line, $0.3M_{\odot}$; solid line, $0.6M_{\odot}$) drawn to further validate the classification of our candidates. As white dwarfs are known to be a hot and blue population, they usually are present at the top left side of the stellar locus in this color-color diagram \citep{eisenstein:06,girven:11}. As expected, the candidates from Set A show a concentrated distribution at the top left, and are well aligned with the expected white dwarf sequence. Stars from Set B as expected show a more dispersed distribution, but the white dwarf candidates still follow the expected white dwarf locus; they also extend somewhat further into the red, which suggests that objects from Set B include cooler white dwarfs on average.

In color-color space, stars from Set B excluded from the white dwarf selection in the RPM diagram (black dots in the top right panel of Figure~\ref{fig:fig8}) mostly fall on the expected locus of main-sequence stars of K-type and early M-type, consistent with our suggestion that they are low-mass stars of the local halo. We found a star among these black dots located in between white dwarfs and giants in the diagram, which implies that it is a binary system with an M dwarf companion \citep{smolvic:04}.  We list our final halo white dwarf candidates from Sets A ($217$ stars) and B ($283$ stars) in Tables~\ref{tab:tab3} and \ref{tab:tab4}, which include \gaia astrometric and photometric measurements and \panstarrs photometric measurements for each one. The description of each column is provided in Table~\ref{tab:tab2}. 

In principle, stars below the red dashed line in the RPM diagram in Figure~\ref{fig:fig8} have the highest likelihood of being local halo white dwarfs, and this line can be used to select halo white dwarfs based on color and reduced proper motion alone, which would be particularly useful for subsets of stars that have no reliable parallaxes. However, we still clearly see a few gray points below the red line, which are stars we excluded from the initial halo selection. A selection based on reduced proper motion alone may thus still be significantly contaminated with thick-disk white dwarfs. \citet{amarante:20} suggested that their model predicts that kinematics of $\sim13$\%~of the high transverse velocity ($v_{T} > 200$ km s$^{-1}$) stars are consistent with that of the thick disk population. this will be an important caveat of our attempt to select halo white dwarfs from Sets C and D.

\subsection{Selection of Halo White Dwarf Candidates from Set C via \panstarrs Photometry}\label{sec:panstarrs_setc}

\begin{figure}[!ht]
\centering
\includegraphics[scale=0.4]{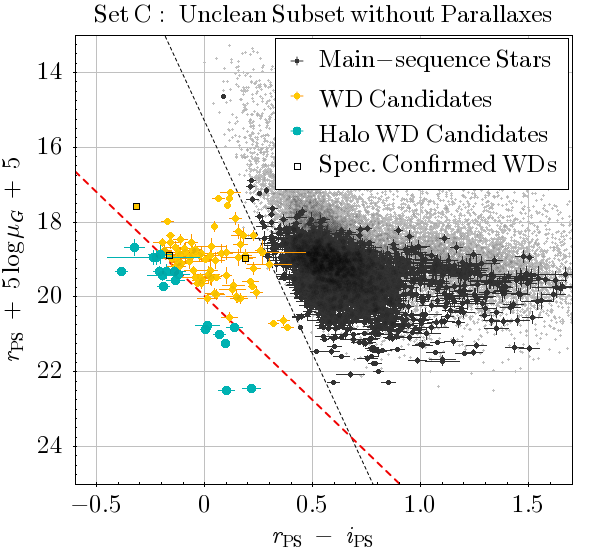}
\includegraphics[scale=0.4]{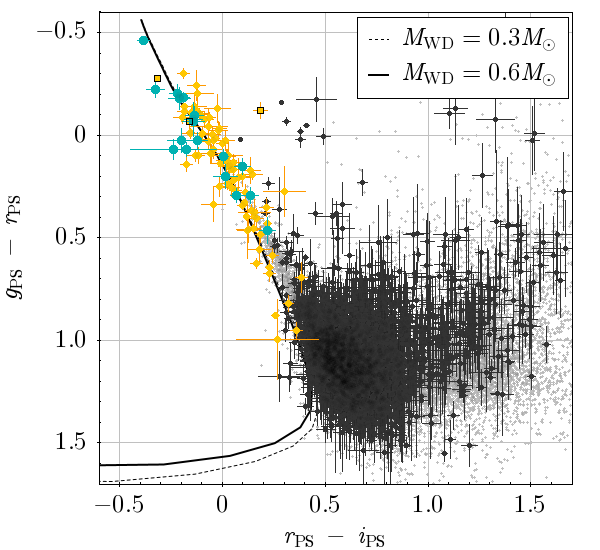}
\caption{Top: RPM diagram built from \gaia proper motion and \panstarrs photometry, for stars in Set C. Stars selected from Figure~\ref{fig:fig3} are overlaid on the distribution, and each symbol separately denotes stars excluded from the selection (black filled circles), $80$ normal white dwarfs (yellow filled circles), and $20$ halo white dwarf candidates (blue filled circles). Error bars show errors in reduced proper motion and \rmi~color. Bottom: \gmr~color distribution as a function of \rmi~color. The lines are the cooling sequences of white dwarfs for pure hydrogen atmospheres: $0.3M_{\odot}$ (dashed line) and $0.6M_{\odot}$ (solid line). Error bars indicate errors in \gmr and \rmi~colors. \label{fig:fig9}}
\end{figure}

We were able to collect \panstarrs photometry for $28{\rm ,}009$ stars in Set C (stars without \gaia parallaxes). Among them, $2571$ stars meet the \gaia RPM diagram selection cut for white dwarfs defined in \S~\ref{sec:rpm_setc}. The top panel in Figure~\ref{fig:fig9} shows the RPM diagram for stars in Set C, based on \panstarrs photometric measurements. We reproduce in this diagram the two selection lines defined in Figure~\ref{fig:fig8} for Sets A and B: the black dashed line separating white dwarfs and main-sequence stars, and the red-dashed line selecting the halo white dwarfs. It is clear that Set C shows massive contamination from main-sequence stars. They are mostly low-mass stars from the clump near the red-dashed line in the \gaia RPM diagram (see Figure~\ref{fig:fig3}). Most of these contaminants (black filled circles) can now be excluded. The remaining $100$ stars located below the black dashed line can still be considered white dwarf candidates (yellow filled circles), and $20$ of them, shown as blue filled circles, are located below the red dashed line, identifying them as possible halo white dwarfs.

The distribution of Set C stars in the color-color plane is shown in the bottom panel in Figure~\ref{fig:fig9}. As expected, stars identified in the RPM diagram as main-sequence stars (black filled circles) have redder colors than normally expected for white dwarfs and are consistent with K or early M dwarfs. On the other hand, the stars identified as white dwarfs based on their RPM diagram distribution form a sequence consistent with the white dwarf locus, and similar to that seen in Figure~\ref{fig:fig8} for the white dwarf candidates from Sets A and B. 

We run the likelihood test for these $100$ white dwarf candidates using the same methods described in \S \ref{sec:setakinematics}. Three conditions are used for the selection: the RPM-\gaia color cut (Equation \ref{eq:eq6}), the RPM-\panstarrs cut (Equation \ref{eq:eq9}), and the RPM-\panstarrs cut for halo white dwarf candidates (Equation \ref{eq:eq10}). The likelihood of being white dwarfs or of being halo white dwarfs is decided by whether a star passes the first two or all three conditions. We confirm that $20$ stars below the red dashed line have likelihood $> 50$\%, making them reasonably likely to be local halo white dwarfs. In Table~\ref{tab:tab5}, we provide the list of $20$ stars with their likelihoods. In addition, we provide information for the remaining $80$ candidates as a reference and because they might include objects of interest.

\subsection{Tentative Identification of Halo White Dwarfs with No \gaia $\bpmrp$ Colors (Set D)}\label{sec:panstarrs_setd}

\begin{figure}
\centering
\includegraphics[scale=0.4]{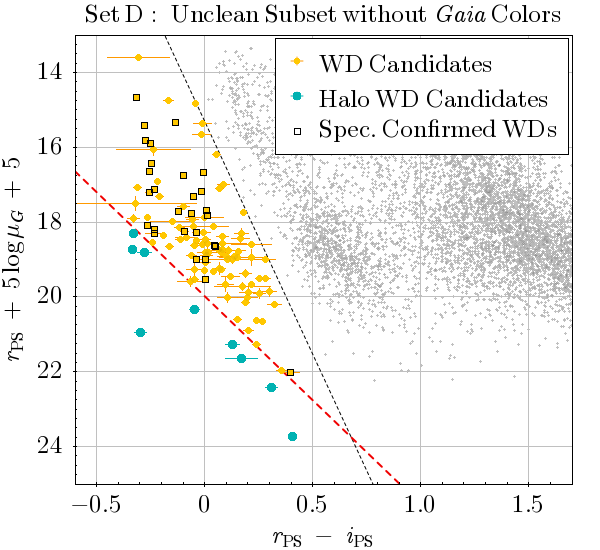}
\includegraphics[scale=0.4]{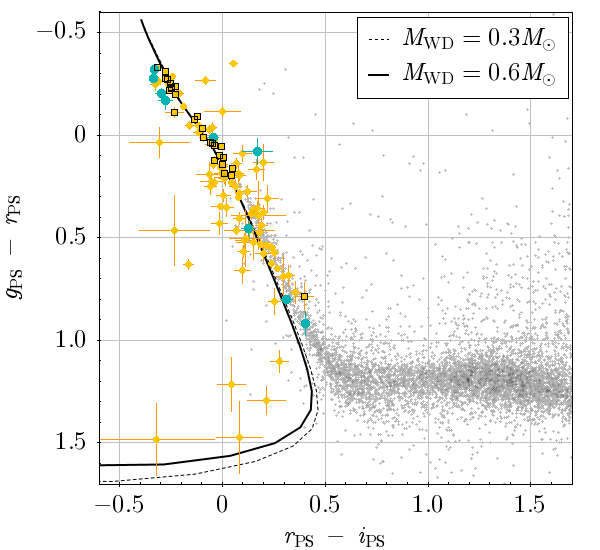}
\caption{Top: same RPM diagram as Figure~\ref{fig:fig10}, but for stars in Set D. Bottom: same color-color diagram as Figure~\ref{fig:fig10}, but for stars in Set D. \label{fig:fig10}}
\end{figure}

Finally, we revisit Set D, which comprises all stars with some astrometric data from \gaia but no $\bpmrp$ colors. Stars in Set D all have proper motions from \gaia, and most also have parallaxes or $G$ magnitudes. Our cross-match method identifies counterparts for $8172$ of these stars in \panstarrs, which allows us to use \panstarrs photometric data instead of \gaia $\bpmrp$~colors. We repeat the analysis we did for Set C to find possible white dwarf candidates from the RPM diagram and color-color diagram. Figure~\ref{fig:fig10} shows the RPM diagram (top panel) and color-color diagram (bottom panel) for all stars in Set D. Applying the same cut, we identify $123$ white dwarf candidates (colored symbols), of which $11$ (blue symbols) met the requirement to be identified as halo white dwarfs.

The RPM diagram for Set D shows a clean separation between the white dwarf and main-sequence loci, similar to that seen for Set A, and it is easy to be convinced that all stars to the left of the dashed line are very likely all white dwarfs. This impression is corroborated by their distribution in the color-color diagram, which has all of the stars falling neatly along the expected white dwarf sequence. 

While most candidates fall onto the cooling sequence of a $0.6 M_{\odot}$~(solid line) white dwarf in the color-color diagram, some candidates appear to be unusually red in \gmr~colors but blue in \rmi. \citet{harris:03} and \citet{kilic:06} reported prominent outliers in the SDSS $(g\,-\,r,\,\,r\,-\,i)$ color-color plane in SDSS, many of which were classified as DC white dwarfs without significant spectral features. They suggested that white dwarfs at cooler temperatures (below $4000$K) are under collision-induced absorption (CIA) due to molecular hydrogen $H_{2}$, which depresses flux in the $i_{\rm PS}$~band and makes a bluer \rmi~color.

In Table~\ref{tab:tab6}, we claim $123$ white dwarf candidates in Set D, and $11$ of them are identified as halo white dwarf candidates with likelihood $> 50$\%. Likelihood values of being white dwarfs or of being halo white dwarfs are provided as well. Note that only nine stars among our local halo candidates are shown in the top panel in Figure~\ref{fig:fig10}, the remaining two stars have extremely blue colors and fall outside the bounds of the plot. This table also includes, for reference, $28$ known stars from Set D identified in various studies, including white dwarf catalogs from SDSS DR7 \citep{debes:11,girven:11,kleinman:13}, SDSS DR10 and DR12 \citep{kepler:15, koester:15}, and LAMOST DR2 \citep{guo:15}.

\begin{figure}
\centering
\includegraphics[scale=0.4]{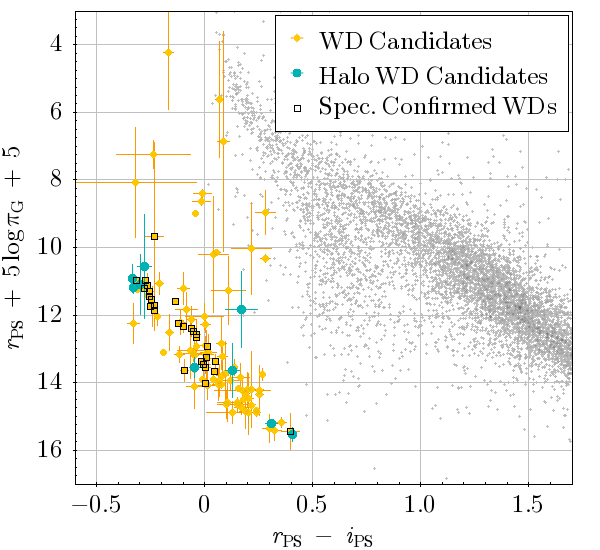}
\caption{CMD for stars in Set D. Yellow points are white dwarf candidates selected from the RPM diagram, and local halo white dwarf candidates are shown as blue points. Overplotted black open boxes represent known white dwarfs identified from various studies (see Table~\ref{tab:tab6}). Error bars indicate errors in absolute magnitude and \rmi~color. \label{fig:fig11}}
\end{figure}

Although Set D is a collection of stars that do not have \gaia colors, \gaia DR2 provides parallaxes for $96$\% of them. Thus, we can use the CMD in order to check the validity of our selection (Figure~\ref{fig:fig11}). We use \rmi~color and an absolute magnitude obtained from the $r_{\rm PS}$ magnitude and the \gaia parallax. The distribution shows both a clear main sequence and a white dwarf cooling sequence, cleaner than the CMD of stars in Set B, which suggests that the \gaia astrometry is useful and reliable, and that Set D is a better subset once it is complemented with external photometric data. Our white dwarf candidates are shown as yellow points, and local halo white dwarf candidates are shown as blue points. Black open boxes show previously known white dwarfs, which mostly fall along the bright end of the white dwarf sequence. While our candidates mostly follow the white dwarf cooling sequence, a few candidates with relatively small error bars in absolute magnitudes and \rmi~colors appear very luminous, which implies that these stars might be unresolved WD+M binaries or hot subdwarfs.

\section{Summary and Conclusions} \label{sec:summary}
With \gaia DR2 we are now having a fresh opportunity to expand the census of local stars of relatively low luminosity, which notably includes the white dwarfs. Expanding the identification and characterization of low-mass stars and white dwarfs of the Galactic halo population, in particular, will fuel further investigations into the star formation history of the local Galactic halo populations, and with greater detail. As one of these first steps, we have identified all white dwarf candidates with large apparent transverse motions ($v_{T} > 200$~km s$^{-1}$) in \gaia DR2 in an area covering $17.3$\% of the sky, based on their kinematics. As a result, we have identified $531$ white dwarfs that are likely to be the remnants of low-mass stars in the local Galactic halo. 

Our candidates were selected from a band with $20^\circ$ width running across both Galactic poles and the Galactic center and anticenter, to facilitate the proper motion selection of objects with large asymmetric drift relative to the local standard of rest. We used \gaia parallaxes, proper motions, and $G$ magnitudes, and we selected stars using a combination of RPM diagram, CMD, and transverse velocities. We divided selected stars into four subsets depending on the quality of their \gaia data: clean subset (A), unclean subset with parallaxes (B), unclean subset without \gaia parallaxes (C), and unclean subset without \gaia colors (D). Since Sets A and B contain parallax measurements, we select halo white dwarf candidates from the projected ($V_{\perp},\,\, V_{\parallel}$) kinematic plane. Our candidates are cross-matched with \panstarrs to obtain more detailed photometric data, not only to better select white dwarf candidates from Sets C and D, but also to confirm our results independent of \gaia photometry. In the case of Sets C and D, we select white dwarf candidates from the RPM diagram as a function of \panstarrs \rmi~color on the basis of an empirical cut defined in the RPM diagram of halo white dwarf candidates from Set A. Although we do not have $griz$ magnitudes for all our stars, because \panstarrs DR1 does not cover our entire sample, the color-color distributions of all subsets in the (\gmr, \rmi) plane confirm that our candidates are mostly white dwarfs, with only minimal contamination from low-mass, main-sequence stars (which are probably also halo members).

To confirm the white dwarf status of our candidates and weed out contaminants, we will ultimately need to collect spectra and do a formal spectroscopic classification. In addition, measuring radial velocities of these objects is required to confirm their kinematic membership (are they actually halo members?) and integrate their Galactic orbit. A full spectroscopic analysis of the subset identified in this paper could also provide masses and cooling ages, which would in turn further constrain the age of the Galactic halo.

\acknowledgments
We are grateful to the anonymous referee for comments that have stimulated significant improvements to the analysis.

This work has made use of data from the European Space Agency (ESA) mission {\it Gaia} (\url{https://www.cosmos.esa.int/gaia}), processed by the {\it Gaia} Data Processing and Analysis Consortium (DPAC, \url{https://www.cosmos.esa.int/web/gaia/dpac/consortium}). Funding for the DPAC has been provided by national institutions, in particular the institutions participating in the {\it Gaia} Multilateral Agreement.

The Pan-STARRS1 Surveys (PS1) and the PS1 public science archive have been made possible through contributions by the Institute for Astronomy, the University of Hawaii, the Pan-STARRS Project Office, the Max-Planck Society and its participating institutes, the Max Planck Institute for Astronomy, Heidelberg, and the Max Planck Institute for Extraterrestrial Physics, Garching, The Johns Hopkins University, Durham University, the University of Edinburgh, the Queen's University Belfast, the Harvard-Smithsonian Center for Astrophysics, the Las Cumbres Observatory Global Telescope Network Incorporated, the National Central University of Taiwan, the Space Telescope Science Institute, the National Aeronautics and Space Administration under Grant No. NNX08AR22G issued through the Planetary Science Division of the NASA Science Mission Directorate, the National Science Foundation Grant No. AST-1238877, the University of Maryland, Eotvos Lorand University (ELTE), the Los Alamos National Laboratory, and the Gordon and Betty Moore Foundation.

Plots were drawn by TOPCAT, which is an interactive graphical software for tabular data \citep{taylor:05}.

\begin{deluxetable*}{ccc}
\tablecolumns{3}
\tablewidth{0pt}
\tabletypesize{\small}
\tablecaption{Assumed Saturation and Detection Limits of \panstarrs Filters \label{tab:tab1}}
\tablehead {
   \colhead{} &
   \colhead{Bright End} &
   \colhead{Faint End} \\
   \colhead{Filter} &
   \colhead{(mag)} &
   \colhead{(mag)}
}
\decimals
\startdata
$g$ & $14.5$ & $22.5$ \\
$r$ & $15.0$ & $22.0$ \\
$i$ & $15.0$ & $21.0$
\enddata
\end{deluxetable*}

\begin{deluxetable*}{lcccchhhhlchhhhcccc}
\tablecolumns{19}
\tablewidth{0pt}
\tabletypesize{\scriptsize}
\tablecaption{High-velocity White Dwarf Candidates in Set A \label{tab:tab3}}
\tablehead {
   \colhead{} &
   \colhead{R.A.} &
   \colhead{Decl.} &
   \colhead{$\pi$} &
   \colhead{$\sigma_{\pi}$} &
   \nocolhead{$\mu_{\alpha}$} &
   \nocolhead{$\mu_{\delta}$} &
   \nocolhead{$G$} &
   \nocolhead{} &
   \colhead{} &
   \colhead{$r_{\rm PS}$} &
   \nocolhead{} &
   \nocolhead{} &
   \nocolhead{$v_{T,\,r}$} &
   \nocolhead{$v_{T,\,s}$} &
   \colhead{$v_{T,\,{\rm tot}}$} &
   \colhead{Halo WD} &
   \colhead{} &
   \colhead{} \\
   \colhead{source\_id} &
   \colhead{(deg)} &
   \colhead{(deg)} & 
   \colhead{(mas)} &
   \colhead{(mas)} &
   \nocolhead{(mas yr$^{-1}$)} &
   \nocolhead{(mas yr$^{-1}$)} &
   \nocolhead{(mag)} &
   \nocolhead{$\bpmrp$} &
   \colhead{PS1\_ID} &
   \colhead{(mag)} &
   \nocolhead{$g_{\rm PS}\,\,-\,\,r_{\rm PS}$} &
   \nocolhead{$r_{\rm PS}\,\,-\,\,i_{\rm PS}$} &
   \nocolhead{(km s$^{-1}$)} &
   \nocolhead{(km s$^{-1}$)} &
   \colhead{(km s$^{-1}$)} &
   \colhead{Likelihood (\%)} &
   \colhead{Spec\_Type} &
   \colhead{References}
}
\decimals
\startdata
  2313554227856823296 & 0.535536 & -32.6280206 & 0.82 & 0.46 & 8.03 & -45.02 & 19.182 & 0.173 & \nodata & \nodata & \nodata & \nodata & 75.48 & -253.3 & 264.3 & 52.14 & \nodata & \nodata  \\
  2313582750735435776 & 0.6362011 & -32.1969737 & 5.131 & 0.124 & 268.55 & -79.49 & 16.36 & -0.317 & \nodata & \nodata & \nodata & \nodata & -188.42 & -177.27 & 258.7 & 100.0 &  DA2  & 6, 8, 29  \\
  2314458438731730304 & 0.8042194 & -31.5123699 & 0.864 & 0.898 & 40.92 & -7.09 & 20.17 & -0.042 & \nodata & \nodata & \nodata & \nodata & -182.72 & -136.19 & 227.89 & 50.04 &  DA  & 11  \\
  2319735617804258176 & 1.7831581 & -31.227056 & 7.721 & 0.11 & 351.93 & -127.25 & 16.665 & -0.263 & \nodata & \nodata & \nodata & \nodata & -155.88 & -168.77 & 229.74 & 100.0 &  DB3  & 4, 6, 13, 29  \\
  4995158325163347712 & 3.4196121 & -42.3757786 & 1.829 & 0.299 & 44.67 & -73.22 & 19.239 & -0.048 & \nodata & \nodata & \nodata & \nodata & -10.57 & -221.97 & 222.22 & 72.76 & \nodata & \nodata  \\
  \ldots              & \ldots     & \ldots      & \ldots & \ldots & \ldots & \ldots & \ldots & \ldots       & \ldots & \ldots & \ldots & \ldots & \ldots & \ldots & \ldots & \ldots \\
  2314173669516064640 & 358.6482487 & -32.355649 & 9.131 & 0.137 & 439.31 & -46.65 & 17.096 & 0.086 & \nodata & \nodata & \nodata & \nodata & -195.04 & -120.61 & 229.32 & 100.0 & WD & 6, 9, 29  \\
\enddata
\tablerefs{(1) \citet{feige:58}, (2) \citet{eggen:65}, (3) \citet{greenstein:76}, (4) \citet{eggen:78}, (5) \citet{green:86}, (6) \citet{mccook:99}, (7) \citet{croom:01}, (8) \citet{koester:01}, (9) \citet{oppenheimer:01}, (10) \citet{lepine:03}, (11) \citet{croom:04}, (12) \citet{kleinman:04}, (13) \citet{salim:04}, (14) \citet{carollo:06}, (15) \citet{eisenstein:06}, (16) \citet{harris:06}, (17) \citet{kilic:06}, (18) \citet{pauli:06}, (19) \citet{kawka:09}, (20) \citet{kilic:10}, (21) \citet{girven:11}, (22) \citet{brown:12}, (23) \citet{kleinman:13}, (24) \citet{li:14}, (25) \citet{kepler:15}, (26) \citet{limoges:15}, (27) \citet{kepler:16}, (28) \citet{leggett:18}, (29) \citet{kilic:19}, (30)\citet{kawka:20}}
\tablecomments{This table is available in its entirety in the machine-readable form.}
\end{deluxetable*}

\begin{deluxetable*}{lcccchhhhlchhhhchhccc}
\tablecolumns{21}
\tablewidth{0pt}
\tabletypesize{\scriptsize}
\tablecaption{High-velocity White Dwarf Candidates in Set B \label{tab:tab4}}
\tablehead {
   \colhead{} &
   \colhead{R.A.} &
   \colhead{Decl.} &
   \colhead{$\pi$} &
   \colhead{$\sigma_{\pi}$} &
   \nocolhead{$\mu_{\alpha}$} &
   \nocolhead{$\mu_{\delta}$} &
   \nocolhead{$G$} &
   \nocolhead{} &
   \colhead{} &
   \colhead{$r_{\rm PS}$} &
   \nocolhead{} &
   \nocolhead{} &
   \nocolhead{$v_{T, \, r}$} &
   \nocolhead{$v_{T, \, s}$} &
   \colhead{$v_{T, \, {\rm tot}}$} &
   \nocolhead{Phot\_Dist} &
   \nocolhead{$v_{T, \, {\rm phot}}$} &
   \colhead{Halo WD} &
   \colhead{} &
   \colhead{}   \\
   \colhead{source\_id} &
   \colhead{(deg)} &
   \colhead{(deg)} & 
   \colhead{(mas)} &
   \colhead{(mas)} &
   \nocolhead{(mas yr$^{-1}$)} &
   \nocolhead{(mas yr$^{-1}$)} &
   \nocolhead{(mag)} &
   \nocolhead{$\bpmrp$} &
   \colhead{PS1\_ID} &
   \colhead{(mag)} &
   \nocolhead{$g_{\rm PS}\, \, -\, \, r_{\rm PS}$} &
   \nocolhead{$r_{\rm PS}\, \, -\, \, i_{\rm PS}$} &
   \nocolhead{(km s$^{-1}$)} &
   \nocolhead{(km s$^{-1}$)} &
   \colhead{(km s$^{-1}$)} &
   \nocolhead{(pc)} &
   \nocolhead{(km s$^{-1}$)} &
   \colhead{Likelihood (\%)} &
   \colhead{Spec\_Type} &
   \colhead{References}
}
\decimals
\startdata 
  2336547386815432960 & 0.4013693 & -24.9199278 & 0.963 & 0.637 & 31.73 & -33.85 & 19.765 & -0.375 & 78090004012856447 & 19.966 & -0.424 & -0.401 & -63.93 & -219.22 & 228.35 & 761.06 & 167.37 & 57.66 & \nodata & \nodata  \\
  2332903880159372672 & 0.4568686 & -29.4269526 & 1.559 & 1.767 & -33.61 & -165.65 & 20.646 & 0.506 & \nodata & \nodata & \nodata & \nodata & 317.77 & -404.01 & 514.01 & 309.78 & 248.19 & 56.85 & \nodata & \nodata  \\
  2417396645764338688 & 0.8341101 & -15.153369 & 5.562 & 1.282 & 99.12 & -242.25 & 20.637 & 0.517 & 89810008340536470 & 20.562 & 0.405 & 0.241 & 22.51 & -221.94 & 223.08 & 303.91 & 377.05 & 66.71 & \nodata & \nodata  \\
  4976547514607879808 & 1.6097728 & -50.158799 & 1.021 & 0.781 & -14.99 & -60.18 & 20.301 & 0.291 & \nodata & \nodata & \nodata & \nodata & 194.0 & -212.57 & 287.79 & 348.34 & 102.4 & 57.97 & \nodata & \nodata  \\
  2417119358380854656 & 2.8410856 & -14.4837046 & 1.633 & 0.654 & 64.74 & -76.99 & 20.046 & -0.255 & \nodata & \nodata & \nodata & \nodata & -55.35 & -286.71 & 292.01 & 612.92 & 292.24 & 78.09 & \nodata & \nodata  \\
  \ldots              & \ldots     & \ldots      & \ldots & \ldots & \ldots & \ldots & \ldots & \ldots       & \ldots & \ldots & \ldots & \ldots & \ldots & \ldots & \ldots & \ldots \\
  2326873849154764416 & 359.4225687 & -29.8951177 & 1.078 & 1.149 & -33.53 & -49.0 & 20.62 & 0.226 & 72123594225446228 & 20.458 & 0.262 & -0.102 & 227.28 & -128.44 & 261.06 & 437.65 & 123.17 & 50.41 & \nodata & \nodata  \\
\enddata
\tablerefs{
(1) \citet{mccook:99}, (2) \citet{croom:01}, (3) \citet{vennes:02}, (4) \citet{lepine:03}, (5) \citet{croom:04}, (6) \citet{kleinman:04}, (7) \citet{eisenstein:06}, (8) \citet{hall:08}, (9) \citet{girven:11}, (10) \citet{west:11}, (11) \citet{carter:13}, (12) \citet{kleinman:13}, (13) \citet{kepler:15}, (14) \citet{fantin:17}, (15) \citet{kilic:19}, (16) \citet{zhang:19}}
\tablecomments{This table is available in its entirety in the machine-readable form.}
\end{deluxetable*}

\begin{deluxetable*}{lcchhhhlchhhcccc}
\tablecolumns{16}
\tablewidth{0pt}
\tabletypesize{\scriptsize}
\tablecaption{White Dwarf Candidates in Set C \label{tab:tab5}}
\tablehead {
   \colhead{} &
   \colhead{R.A.} &
   \colhead{Decl.} &
   \nocolhead{$\mu_{\alpha}$} &
   \nocolhead{$\mu_{\delta}$} &
   \nocolhead{$G$} &
   \nocolhead{} &
   \colhead{} &
   \colhead{$r_{\rm PS}$ } &
   \nocolhead{} &
   \nocolhead{} &
   \nocolhead{} &
   \colhead{WD} &
   \colhead{Halo WD} &
   \colhead{} &
   \colhead{}  \\
   \colhead{source\_id} &
   \colhead{(deg)} &
   \colhead{(deg)} & 
   \nocolhead{(mas yr$^{-1}$)} &
   \nocolhead{(mas yr$^{-1}$)} &
   \nocolhead{(mag)} &
   \nocolhead{$\bpmrp$} &
   \colhead{PS1\_ID} &
   \colhead{(mag)} &
   \nocolhead{$g_{\rm PS}\, \, -\, \, r_{\rm PS}$} &
   \nocolhead{$r_{\rm PS}\, \, -\, \, i_{\rm PS}$} &
   \nocolhead{Halo\_Status} &
   \colhead{Likelihood (\%)} &
   \colhead{Likelihood (\%)} &
   \colhead{Spec\_Type} &
   \colhead{References}
}
\decimals
\startdata
  2417140047238388864 & 3.6801333 & -14.3027568 & 95.22 & -16.06 & 19.608 & 0.673 & 90830036799957120 & 19.544 & 0.219 & 0.054 & \nodata & 100.0 & 0.0 & \nodata & \nodata \\
  2320117629375536512 & 5.1560428 & -29.920615 & 45.73 & 3.26 & 20.49 & 0.243 & 72090051559605579 & 20.478 & 0.07 & -0.041 & \nodata & 100.0 & 0.0 & \nodata & \nodata \\
  2349743076832651648 & 12.5277632 & -20.9737167 & 58.39 & -10.95 & 19.396 & 0.151 & 82830125277821944 & 19.126 & 0.145 & -0.172 & \nodata & 100.0 & 0.0 & \nodata & \nodata \\
  2355740496149824640 & 12.918436 & -20.8993984 & 47.17 & -3.23 & 20.335 & 0.229 & 82920129184271074 & 20.339 & -0.008 & -0.095 & \nodata & 100.0 & 0.01 & \nodata & \nodata \\
  2355806024466166784 & 14.0156541 & -20.6681025 & 46.74 & -21.32 & 20.891 & 0.341 & 83190140156208650 & 20.837 & 0.026 & -0.119 & * & 100.0 & 59.57 & \nodata & \nodata \\
  \ldots              & \ldots     & \ldots      & \ldots & \ldots & \ldots & \ldots & \ldots & \ldots       & \ldots & \ldots & \ldots & \ldots & \ldots & \ldots & \ldots \\
  2339318877674994816 & 358.0884434 & -22.9675101 & -21.92 & -40.33 & 20.543 & 0.957 & 80433580884319341 & 20.502 & 0.202 & 0.099 & \nodata & 85.71 & 0.0 & \nodata & \nodata \\
\enddata
\tablerefs{(1) \citet{eisenstein:06}, (2) \citet{kleinman:13}}
\tablecomments{This table is available in its entirety in the machine-readable form.}
\end{deluxetable*}

\begin{deluxetable*}{lcccchhhlchhhcccc}
\tablecolumns{17}
\tablewidth{0pt}
\tabletypesize{\scriptsize}
\tablecaption{White Dwarf Candidates in Set D \label{tab:tab6}}
\tablehead {
   \colhead{} &
   \colhead{R.A.} &
   \colhead{Decl.} &
   \colhead{$\pi$} &
   \colhead{$\sigma_{\pi}$} &
   \nocolhead{$\mu_{\alpha}$} &
   \nocolhead{$\mu_{\delta}$} &
   \nocolhead{$G$} &
   \colhead{} &
   \colhead{$r_{\rm PS}$} &
   \nocolhead{} &
   \nocolhead{} &
   \nocolhead{} &
   \colhead{WD} &
   \colhead{Halo WD} &
   \colhead{} &
   \colhead{}  \\
   \colhead{source\_id} &
   \colhead{(deg)} &
   \colhead{(deg)} & 
   \colhead{(mas)} &
   \colhead{(mas)} &
   \nocolhead{(mas yr$^{-1}$)} &
   \nocolhead{(mas yr$^{-1}$)} &
   \nocolhead{(mag)} &
   \colhead{PS1\_ID} &
   \colhead{(mag)} &
   \nocolhead{$g_{\rm PS}\,\,-\,\,r_{\rm PS}$} &
   \nocolhead{$r_{\rm PS}\,\,-\,\,i_{\rm PS}$} &
   \nocolhead{Halo\_Status} &
   \colhead{Likelihood (\%)} &
   \colhead{Likelihood (\%)} &
   \colhead{Spec\_Type} &
   \colhead{References}
}
\decimals
\startdata
  2361704045355924352 & 5.2793135 & -21.0576421 & 4.355 & 0.222 & -34.45 & -34.98 & 17.285 & 82730052791791209 & 15.295 & 1.193 & -1.324 & * & 100.0 & 100.0 & \nodata & \nodata \\
  5041666743597100928 & 19.655093 & -22.9164157 & 5.461 & 0.175 & 31.32 & -82.7 & 17.677 & 80500196549740655 & 16.453 & -0.349 & 0.056 & \nodata & 80.84 & 0.0 & \nodata & \nodata \\
  2482500275433240832 & 21.0240977 & -4.3160092 & 1.543 & 0.713 & 24.87 & -41.54 & 20.375 & 102820210240741366 & 20.326 & 0.504 & 0.11 & \nodata & 98.09 & 0.0 & \nodata & \nodata \\
  5121322444019559936 & 32.8875957 & -24.837461 & 6.226 & 1.632 & 19.43 & -72.89 & 20.797 & 78190328874785551 & 20.631 & 0.515 & 0.105 & \nodata & 100.0 & 2.6 & \nodata & \nodata \\
  5075663260876120064 & 42.2124742 & -26.5199788 & 5.141 & 0.97 & 1.89 & -40.74 & 20.626 & 76170422124306424 & 20.515 & 0.209 & 0.014 & \nodata & 100.0 & 0.0 & \nodata & \nodata \\
  \ldots              & \ldots     & \ldots      & \ldots & \ldots & \ldots & \ldots & \ldots & \ldots       & \ldots & \ldots & \ldots & \ldots & \ldots & \ldots & \ldots & \ldots \\
  1047465080939412608 & 154.6366483 & 59.0896157 & -0.269 & 0.68 & -53.88 & -39.14 & 19.771 & 178901546367758392 & 18.875 & 2.083 & -0.148 & \nodata & 99.98 & 1.93 & \nodata & \nodata \\
\enddata
\tablerefs{(1) \citet{mccook:99}, (2) \citet{debes:11}, (3) \citet{girven:11}, (4) \citet{kleinman:13}, (5) \citet{guo:15}, (6) \citet{kepler:15}, (7) \citet{koester:15}, (8) \citet{limoges:15}, (9) \citet{dame:16}, (10) \citet{kepler:16} }
\tablecomments{This table is available in its entirety in the machine-readable form.}
\end{deluxetable*}

\begin{deluxetable*}{cl}
\tablecolumns{2}
\tablewidth{0pt}
\tabletypesize{\footnotesize}
\tablecaption{Column Description Provided in Tables\label{tab:tab2}}
\tablehead{
	\colhead{Header} &
	\colhead{Description}
}
\startdata
 source\_id & \gaia DR2 unique source identifier \\
 RA & Right ascension in J2015.5 (deg) \\
 Dec & Declination in J2015.5 (deg)  \\
 $\pi$ & \gaia DR2 parallax (mas) \\
 $\sigma_{\pi}$ & \gaia DR2 standard error of parallax (mas) \\
 $\mu_{\alpha}$ & \gaia proper motion in right ascension direction (mas yr$^{-1}$) \\
 $\mu_{\delta}$ & \gaia proper motion in declination direction (mas yr$^{-1}$) \\
 $G$ & \gaia $G$-band mean magnitude (mag) \\
 $\bpmrp$ & \gaia BP - RP color (mag) \\
 PS1\_id & \panstarrs DR1 identifier \\
 $r_{\rm PS}$ & \panstarrs $r$-band magnitude (mag) \\
 \gmr & \panstarrs \gmr~color (mag) \\
 \rmi & \panstarrs \rmi~color (mag) \\
 $v_{T,\,r}$ & Transverse velocity in $r$~direction from \gaia parallax (km s$^{-1}$) \\
 $v_{T,\,s}$ & Transverse velocity in $s$~direction from \gaia parallax (km s$^{-1}$) \\
 $v_{T,\,{\rm tot}}$ & Total transverse velocity from \gaia parallax (km s$^{-1}$)\\
 Phot\_Dist & Photometric distance for Set B only (pc) \\
 $v_{T,\,{\rm phot}}$ & Total transverse velocity from photometric distance for Set B only (km s$^{-1}$)\\
 WD Likelihood & Confidence level to pass the white dwarf selection (\%)\\
 Halo WD Likelihood & Confidence level to pass the the halo white dwarf selection expressed (\%)\\
 Halo\_Status & Halo status of white dwarf candidates for Sets C and D only\tablenotemark{a} \\
 Spec\_Type & Spectral type of a known white dwarf given in the SIMBAD database \\
 Ref. & References \\
\enddata
\tablenotetext{a}{\*: Halo white dwarf candidates with likelihood $> 50$\%}
\end{deluxetable*}
 
\end{document}